\documentclass[journal]{IEEEtran}
\usepackage[cmex10]{amsmath}
\usepackage{graphicx}
\usepackage{algorithm}
\usepackage{algpseudocode}
\usepackage{amsthm}
\theoremstyle{remark}
\newtheorem*{remark}{Remark}
\usepackage{booktabs}
\usepackage{multirow}
\usepackage{amsfonts}
\usepackage{url}
\usepackage{diagbox}
\usepackage{hhline}
\usepackage{enumitem}
\usepackage[noadjust]{cite}
\usepackage{epstopdf}
\usepackage[usenames,dvipsnames,svgnames,table]{xcolor}
\ifCLASSINFOpdf
\else
\fi

\begin{document}

\title{Decentralized Stochastic Optimal Power Flow\\ in Radial Networks with Distributed Generation}

\author{Mohammadhafez Bazrafshan,~\IEEEmembership{Student Member,~IEEE}, 
       and Nikolaos Gatsis,~\IEEEmembership{Member,~IEEE}
\thanks{Manuscript received February 9, 2015; revised June 12, 2015 and October 30, 2015; accepted January 4, 2016.} \thanks{The authors are with the department of electrical and computer engineering, the University of Texas at San Antonio, San Antonio, TX 78249, USA (emails: aju084@my.utsa.edu, nikolaos.gatsis@utsa.edu)}
  \thanks{This material is based upon work supported by the National Science Foundation under Grant No. CCF-1421583.}}

\markboth{IEEE Transactions on Smart Grid}%
{}

\maketitle

\begin{abstract}
This paper develops a power management scheme that jointly optimizes the real power consumption of programmable loads and reactive power outputs of photovoltaic (PV) inverters in distribution networks.  The premise is to determine the optimal demand response schedule that accounts for the stochastic availability of solar power, as well as to control the reactive power generation or consumption of PV inverters \textit{adaptively} to the real power injections of all PV units.  These uncertain real power injections by PV units are modeled as random variables taking values from a finite number of possible scenarios. Through the use of second order cone relaxation of the power flow equations, a convex stochastic program is formulated. The objectives are to minimize the negative user utility, cost of power provision, and thermal losses, while constraining voltages to remain within specified levels. To find the global optimum point, a decentralized algorithm is developed via the alternating direction method of multipliers that results in closed-form updates per node and per scenario, rendering it suitable to implement in distribution networks with large number of scenarios. Numerical tests and comparisons with an alternative deterministic approach are provided for typical residential distribution networks that confirm the efficiency of the algorithm.  
\end{abstract}


\IEEEpeerreviewmaketitle

\section{Introduction}

\IEEEPARstart{R}{esidential}-scale solar photovoltaic (PV) systems are paving their way into today's distribution systems, affecting higher incorporation of distributed generation into modern power systems.  The chief advantage is that energy is generated closer to the point of consumption, thereby helping to reduce the transmission network congestion.  A major challenge in incorporating PV systems in distribution networks is the uncertain availability of solar energy due to changes in irradiance conditions.  These changes lead to insufficient or at times excess electricity generation, and if unaccounted for, can result in reduced user satisfaction, poor voltage regulation, and eventually equipment failure.  

To deal with these issues, programmable loads that enable control of their real power consumption provide an opportunity for distribution system operators (DSO) to reduce the peak load in periods of inadequate generation. Moreover, reactive power generation or consumption by PV inverters, which provide the AC interface between the PV system and the grid, can be leveraged to improve voltage regulation.   Although current standards prohibit these inverters to operate at a variable power factor \cite{ieee1547}, the potential advantages of using these capabilities for voltage regulation have been extensively reported in literature; see e.g., \cite{TuSuBaCh-ProcIEEE11, baran2007, adaptivelearning}. 

Building upon the aforementioned capabilities, this paper proposes a decentralized real and reactive power management framework in distribution systems with high levels of PV generation.  To account for the inherent uncertainty in solar power generation, stochastic programming tools \cite{Conejo-Uncert} are used to achieve common objectives such as loss minimization, operational cost reduction, and acceptable voltage regulation.

\subsection{Prior Art} 
Power management in distribution networks amounts to an optimal power flow  (OPF) problem that minimizes certain objectives subject to power flow equations that are generally nonconvex.  See \cite{BaWu89, baranWu3 , bwu89LoadBalance},  for the canonical form of power flow equations in radial distribution networks.   Due to the nonconvexity of power flow equations, many relaxations and approximations have been recently proposed.   A comprehensive study summarizing the recent advances in convex relaxations of OPF can be found in \cite{low-co-ex} and \cite{low-co}. In particular, conic relaxation techniques are used in \cite{Jabr2006} for radial distribution load flow, while \cite{Lavaei2012}, \cite{SoLa2012}, and \cite{GaLiToLo2015} provide conditions to guarantee optimality of relaxations to the original OPF. 

Deterministic approaches to reactive power management in distribution networks have recently been investigated in many studies where user power consumption and PV power generation are known.  The reactive power management problem is approached in \cite{TSBC10a}, \cite{TSBC10b}, and \cite{TuSuBaCh-ProcIEEE11} using a linear approximation of the power flow equations, called \texttt{LinDistFlow} equations, and local reactive power control policies. Although these local policies are computationally attractive and perform well in practical scenarios, they do not provide optimality guarantees.  

An optimal decentralized algorithm for solving the reactive power control problem under the \texttt{LinDistFlow} model is developed in \cite{SuBaCh-arx13a}  using the alternating direction method of multipliers (ADMM).   An adaptive VAR control scheme is pursued in \cite{YGL12} based on the \texttt{LinDistFlow} where the adaptation law switches between minimizing power losses or maintaining voltage regulation.   Centralized reactive power control and PV inverter loss minimization using the second-order cone programming (SOCP) relaxation is the theme of \cite{FNCL12}.  

Decentralized solvers for real and reactive power optimization using the SOCP relaxations are developed in \cite{LiGaChLo12} and \cite{LiChLo12} using the Predictor Corrector Proximal Method of Multipliers (PCPM).   Leveraging the SOCP relaxation  and the ADMM, a decentralized solver for OPF with closed-form updates is designed in \cite{PeLo14} where the user-consumed reactive power is modeled as independent of users' real power consumption.  Decentralized real power control using ADMM with convex envelop approximations is developed in \cite{KrChLaBo13}.  Leveraging semidefinite programming (SDP) relaxations, decentralized algorithms are designed in \cite{DaZhGi14} and \cite{DaDhJoGi14} using ADMM, and in \cite{LaZhDoTs12} via a dual subgradient method.  Distributed reactive power control is performed in \cite{Robbins13} with the purpose of maintaining nodal voltages within specification limits. 

A nonconvex formulation for the reactive power control problem is presented in \cite{voltvar2012}, and a solver based on sequential convex programming is developed, but without global optimality guarantees. An adaptive local learning algorithm is devised in \cite{adaptivelearning} which provides a fast approximate solution for voltage regulation  using the solutions of past optimizations. 

Up to this point, all previously mentioned approaches assume that real power injections in buses with renewable generation are known, and hence are deterministic.  The works in \cite{distcont} and \cite{bolognani2015} propose distributed  online algorithms for optimal reactive power compensation and loss minimization based on feedback control from local voltage measurements.  By modeling local voltages as functions of reactive power in microgenerators,  the loss terms are casted as a quadratic form in reactive powers, and a modified dual method is used to minimize the losses subject to power constraints. These works model the real power injection of distributed generators as unmeasured disturbances.

In this light, the work in \cite{nali2014conf} also assumes no knowledge of the real power injection of distributed generators. It is shown that a purely local reactive power control that leverages only voltage measurements and does not rely on communication, does not by itself guarantee acceptable voltage regulation.  Therefore, to perform voltage regulation and provide optimality guarantees, additional information,  such as previous control inputs is incorporated in their proposed algorithm.

 The work in \cite{conejo-giannakis14} models loads and real power injections on nodes as stochastic processes, collects noisy and delayed estimates of those, and decides the reactive power injection by solving a centralized optimization problem using a stochastic approximation algorithm. Uncertainty-aware optimal real and reactive power management from PV units is analyzed in \cite{DaDhJoGi15} where the conditional-value-at-risk is utilized to minimize the risk of overvoltages.

\subsection{Contributions and outline}
The contributions of this paper are as follows: 
\begin{enumerate}
\item An optimal power flow problem for distribution networks accounting for the uncertainty in solar generation is formulated. 
In particular, the real powers generated by PV units at different nodes are modeled as random variables that take values from a finite set of scenarios. Real power of user controllable loads is optimized jointly with reactive power injection or absorption by PV inverters and network power flows per scenario. This stochastic model captures the uncertainty in solar generation which is not accounted for in previous approaches \cite{TuSuBaCh-ProcIEEE11}--\cite{adaptivelearning} and \cite{TSBC10a}-\cite{voltvar2012}, which assume known (deterministic) PV injections.  In contrast to \cite{distcont}--\cite{nali2014conf}, in which voltage regulation through reactive power control is based on feedback from local voltage measurements, our methodology maintains voltages within specified limits while accounting for the underlying uncertainty of the PV generation.  Recently, uncertainty in distributed PV generation has been addressed in~\cite{DaDhJoGi15} and~\cite{conejo-giannakis14}. Reactive power control by PV units and power flows are decided in the present work adaptively to solar power outputs, as opposed to a static fashion in~\cite{DaDhJoGi15}. User controllable loads are not modeled in~\cite{conejo-giannakis14}, which in addition features a centralized optimization scheme, in contrast to the decentralized solution algorithm developed here (featured as the second contribution next).
\item This paper develops a decentralized solver with the following desirable attributes, motivated by scalability considerations: (a1) updates are decomposed per node and per scenario; (a2) all updates are in closed-form and (a3) communication only between neighboring nodes is required. The decentralized algorithm is based on ADMM. Even though ADMM naturally lends itself to distributed computation, there are two challenges that need to be addressed, in order to successfully arrive to an optimization algorithm with the previously mentioned attributes: (c1) introduce properly designed auxiliary variables, and (c2) appropriately split the set of constraints into coupling and individual constraints. Different choices for (c1) and (c2) may lead to entirely different algorithms; one of this paper's contributions is to address those challenges towards a fully decentralized solver with closed-form updates. The decentralization methodology
in this paper is extending the ADMM approach in \cite{PeLo14} in a stochastic programming setup. In addition, the closed-form solution of an SOCP in 4 variables with upper bounds on certain variables is derived, motivated by the incorporation of line current limits in the present formulation.
This paper also expands previous work in \cite{BG-Sg2014}---which featured only single-line networks and the \texttt{LinDistFlow} approximations---in two major ways: 
(1) SOCP relaxation of power flow equations is incorporated, which is a more complex but also more accurate model; and (2) the formulation incorporates tree networks. 
\item Considering realistic PV generation models and only a small number of representative scenarios for the uncertainty, numerical tests highlight the benefits of the proposed stochastic formulation with regards to user satisfaction and thermal loss minimization. Improved voltage regulation and reduced thermal losses are demonstrated in comparison to alternative distributed control schemes in~\cite{TuSuBaCh-ProcIEEE11}, and also in networks that include shunt capacitors.
\end{enumerate} 
The remainder of this paper is organized as follows. Section~\ref{sec:netmod} introduces the network model, the decision variables, and pertinent constraints. The optimization problem is formulated in Section~\ref{sec:opf}. Section~\ref{sec:algorithm} develops the equivalent formulation suitable for decentralized solution by the ADMM, and derives the closed-form updates. Section \ref{sec:commreq} deliberates on the algorithm implementation and the communication requirements. Numerical tests are provided in Section~\ref{sec:numtest} as well as comparisons with competing approaches. Section~\ref{sec:conc} concludes the paper. 

\section{Network Model and Decision Variables}
\label{sec:netmod}

\par
\begin{figure}[t]
\centering
\includegraphics[scale=0.35]{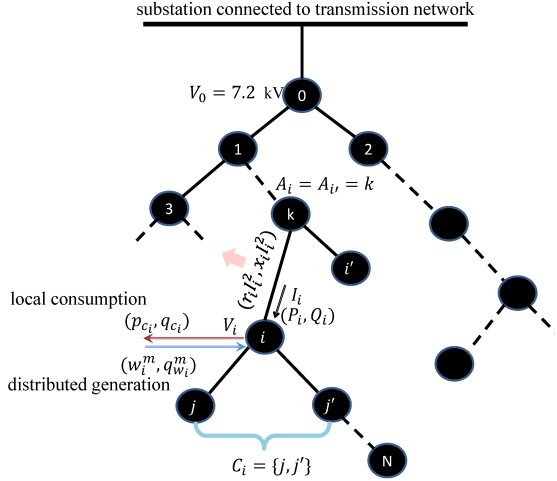}
\caption{A radial distribution network modeled as a tree graph.}
\label{fig:trne}
\end{figure}

{}Consider a radial distribution network as depicted in Fig.~\ref{fig:trne} modeled by a tree graph, where the set of all nodes is denoted as $\mathcal{N}=\{0,1,2,...,N\}$.  Node $0$ (i.e., root of the graph) is the substation connected to the transmission network, and the remaining  $N$ nodes represent users.

Following the tree model for distribution networks, each node $i \in \mathcal{N} \setminus \{0\}$ has a unique ancestor denoted by $A_i$. The line connecting node $A_i$ to node $i$ is labeled as line $i$, and is considered to have resistance $r_i$ and reactance $x_i$.  Each node $i \in \mathcal{N}$ has an associated set of child nodes denoted by $\mathcal{C}_i$.  For the terminal nodes (i.e., leaves of the tree, or nodes without children), it holds that $\mathcal{C}_i=\emptyset$. 
\subsection{User load model}
User $i$ consumes a non-elastic real and reactive load denoted by $P_{L_i}$ and $Q_{L_i}$ respectively.  Moreover, users are supposed to have demand response capabilities, and their elastic consumption $p_{c_i}$ is permitted to vary in a certain range:
\begin{align}
\label{eqn:rebo}
0 \le p_{c_i} \le p_{c_i}^{\max}, \quad i \in \mathcal{N} \setminus \{0\}.
\end{align}
The elastic reactive power consumption $q_{c_i}$ has a linear relationship with the real power $p_{c_i}$:
\begin{align}
\label{eqn:pf}
q_{c_i}=\left(\sqrt{\frac{1}{\mathrm{PF}_i^2}-1}\right)~p_{c_i}, \quad i \in \mathcal{N} \setminus \{0\}.
\end{align}
where  $\mathrm{PF}_i$ is the power factor,  a dimensionless number in the interval $(0,1]$.

\subsection{PV generation model}
\label{sub:stochmodel}
User nodes may also be enabled with PV generation. Attributable to the stochastic nature of solar power, the real power injections of PV systems are modeled as random variables.  The real power injections across the network take values from a finite set of $M$ possible scenarios, each with probability $\pi^m$, with $m \in \mathcal{M}= \{1,2,\ldots,M\}$.  It is thus assumed that the real power generated by the PV unit at node $i$ and in scenario $m$ is given by $w_i^m$.  A typical probabilistic model for generating scenarios is the beta distribution, see e.g., \cite{WaChWaBe2014} and \cite{NiZaAg2012}. The mean value of the distribution can be set to a forecasted generation for the next time period that could range from e.g., 15 minutes to 1 hour.

The DC electrical output generated by the PV modules are translated into an AC output through the use of PV inverters.  These PV inverters are also capable of generating or consuming reactive power  by themselves; see e.g., \cite{TuSuBaCh-ProcIEEE11}.   Let $q_{w_i}^m$ denote the  reactive power generated by the PV inverter at node $i$ in scenario $m$.  Then, $q_{w_i}^m$ is a decision variable constrained by
\begin{align}
\label{eqn:reabo}
-q_{w_i}^{\max} \le q_{w_i}^m \le q_{w_i}^{\max}, \quad  i \in \mathcal{N} \setminus \{0\}, m \in \mathcal{M}
\end{align} 
where $q_{w_i}^{\max}=\sqrt{s_{w_i}^2-(w_i^m)^2}$ and $s_{w_i}$ is the maximum apparent power capacity of the PV at node $i$, that is, the nameplate capacity of the PV inverter at that node.

\subsection{Power flow equations}
The scenario-dependency of real and reactive power injections of PV units renders the power flow equations across the network to be scenario dependent as well. 
 At scenario $m$, real and reactive power flows on line $i$ are denoted by $P_i^m$ and $Q_i^m$ respectively; the squared magnitude of the voltage phasor at node $i$  and the squared magnitude of the current phasor on line $i$ are represented by $v_i^m$ and $l_i^m$, respectively. 
The power flow equations leveraging the SOCP relaxation are as follows, where all the constraints hold for  $m\in\mathcal{M}$ and $i\in \mathcal{N}$:
\begin{IEEEeqnarray}{rCl}
P_i^m &=& \sum\limits_{j\in \mathcal{C}_i}(P_j^m+r_jl_j^m)+P_{L_i}+p_{c_i}-w_i^m   \label{eqn:repofl}\\ 
Q_i^m &=& \sum\limits_{j \in \mathcal{C}_i}(Q_j^m+x_jl_j^m)+Q_{L_i}+q_{c_i}-q_{w_i}^m-q_{s_i}v_i^m\label{eqn:reapofl}\\
v_{A_i}^m&=&v_{i}^m+2(r_iP_i^m+x_iQ_i^m)+(r_i^2+x_i^2)l_i^m\label{eqn:vopofl}, i \neq 0, 
\end{IEEEeqnarray}
\begin{IEEEeqnarray}{rCl}
(P_i^m)^2+(Q_i^m)^2 &\le& v_i^ml_i^m,  i \neq 0,\label{eqn:sopofl}\\
v_i^m &\ge& 0, i \neq 0.  \label{eqn:vocupofl} 
\end{IEEEeqnarray}

In the equations above, $r_i$ and $x_i$ have units of $\Omega$.  Real powers have units of $\mathrm{MW}$ and reactive powers are in $\mathrm{MVars}$.  Square currents are in $(kA)^2$ while square magnitude of voltages $v_i^m$ have units of $(\mathrm{kV})^2$.  Note that $v_i^m=(V_i^m)^2$ where $V_i^m$ is the magnitude of the voltage phasor at node $i$ and scenario $m$.  The substation voltage is fixed at $v_0=v_{0}^m=(V_0^m)^2$ and  since there is no user at the substation, $p_{c_0}, w_0^m, q_{w_0}^m$ are all zero.

The power flow equations \eqref{eqn:repofl}-\eqref{eqn:vocupofl} clearly show that fluctuations in solar real power injection ($w_i^m$) ultimately lead to variations in voltage levels ($v_i^m$) across nodes in the network.  In order to  guarantee that node voltages remain within safety levels, the following voltage regulation constraint is enforced at every node $i \in \mathcal{N} \setminus \{0\}$ per scenario $m \in \mathcal{M}$
\begin{IEEEeqnarray}{rCl}
\label{eqn:vore}
(1-\epsilon)^2 &\le& \frac{v_i^m}{v_0} \le (1+\epsilon)^2 
\end{IEEEeqnarray}
where $\epsilon$ could be chosen to be 0.05. The current magnitudes for every line are capped as per the following constraint:
\begin{IEEEeqnarray}{rCl}
0 &\le & l_i^m \le l_i^{\max}. \label{eqn:linelimit}
\end{IEEEeqnarray}
A summary of decision variables along with the user and network parameters is given in Table~\ref{table:variabledescription}. 
\begin{table}
\caption{Decision variables,  user and network parameters}
\label{table:variabledescription}
\centering
\renewcommand{\arraystretch}{1.5} 
\begin{tabular}{|c|c|}
\hline
\multicolumn{2}{|c|}{Decision variables}   \\
\hline
Scenario-independent & $p_{c_i}$, $q_{c_i}$ \\
\hline
Scenario-dependent & $P_i^m$, $Q_i^m$, $v_i^m$, $l_i^m$, $q_{w_i}^m$ \\
\hline
\hline
\multicolumn{2}{|c|}{Parameters (known)}  \\
\hline
User  & $P_{L_i}$, $Q_{L_i}$, $\mathrm{PF}_i$, $p_{c_i}^{\max}$, $w_i^m$, $s_{w_i}$ \\
\hline
Network  & $r_i$, $x_i$, $q_{s_i}$, $\epsilon$, $l_i^{\max}$\\
\hline
\end{tabular}
\end{table}

Having described the optimization variables and constraints, the next section elaborates on the relevant objective function and completes the problem formulation.
\section{Optimal power flow formulation}
\label{sec:opf}
One of the main objectives for a DSO is to meet customer demand.  In order to quantify user satisfaction for demand response decisions,  a concave utility function denoted as $u_i(p_{c_i})$ is adopted for each user $i$.   Maximizing the sum of user utilities hence constitutes the first objective.  

Power flows in the network are provided through the substation that is connected to the transmission grid.  The DSO undergoes a cost to obtain this power from the transmission network.  Although any cost function $C(P_0^m)$ that is convex can be used to evaluate the cost of power provision at scenario $m$, a particular cost function of interest could be one that distinguishes between the buying price (import) and the selling price (export).  One such example is a piece-wise linear function of the form:
\begin{equation}
\label{eqn:coP0}
C(P_0^m)=
\left\{
	\begin{array}{ll}
		aP_0^m  & \mbox{if } P_0^m \geq 0 \\
		bP_0^m & \mbox{if } P_0^m < 0
	\end{array}
\right.
\end{equation} 
with $a > b \ge 0$ to preserve convexity. 
The expected value of $C(P_0^m)$ over all scenarios, that is $\sum\limits_{m=1}^M \pi^mC(P_0^m)$ is the second term in the objective.
The third term in the objective is the expected incurred thermal losses in the lines over all the scenarios, that is $\sum\limits_{m=1}^M\pi^m\sum\limits_{i=1}^Nr_il_i^m$.   

Let $\mathbf{P}, \mathbf{Q}, \mathbf{v}, \mathbf{l}, \mathbf{p_{c}}, \mathbf{q_w}$ collect the respective variables per node and per scenario (if the variable is scenario dependent). The optimization problem amounts to 
\begin{IEEEeqnarray}{rl}
\label{eqn:mainQP}
\min_{\substack{\mathbf{P}, \mathbf{Q}, \mathbf{v} \\ \mathbf{l}, \mathbf{p_{c}}, \mathbf{q_w}}}~ ~& -\sum\limits_{i=1}^N u_i(p_{c_i}) + \sum\limits_{m=1}^M \pi^m C(P_0^m) \nonumber \\ &+K_\mathrm{Loss}\sum\limits_{m=1}^M\pi^m\sum\limits_{i=1}^Nr_il_i^m \\
\text{subject to }& ~  \eqref{eqn:rebo}-\eqref{eqn:linelimit} \notag
\end{IEEEeqnarray}
where  $K_{\mathrm{Loss}} \ge 0$ is a weight which can be selected by the system operator to reflect the relative priority of loss minimization with respect to the two other objective terms.

Problem \eqref{eqn:mainQP} forms a two-stage stochastic convex program with first-and-second-stage decisions.  First-stage decisions are determined independently of the uncertainty and comprise elastic load consumptions $p_{c_i}$'s (and ultimately $q_{c_i}$'s). Second-stage decisions are determined adaptively to the uncertainty and include reactive power injection/absorption provided by the PV inverters ($q_{w_i}^m$),  power flows  ($P_i^m, Q_i^m$), and squared magnitude of voltages and currents ($v_i^m$ and $l_i^m$ for every scenario).
\begin{remark}[Compatibility of the formulation with any convex cost function]
Auxiliary variables can be used to relieve the difficulty of directly working with piecewise linear (nondifferentiable) functions such as \eqref{eqn:coP0} in the objective. For example, $C(P_0^m)$ can be written as
\begin{align}
\label{eqn:specialcost}
C(P_0^m)=aP_{0+}^m-bP_{0-}^m 
\end{align} 
and the following constraints are added (for $m \in \mathcal{M}$):
\begin{align}
\label{eqn:P0originalconstr}
P_0^m=P_{0+}^m-P_{0-}^m, \quad  P_{0+}^m \ge 0 , P_{0-}^m\ge 0  .
\end{align}

Solving \eqref{eqn:mainQP} with $C(P_0^m)$ given by \eqref{eqn:specialcost} and with added constraints of \eqref{eqn:P0originalconstr} is equivalent to solving \eqref{eqn:mainQP} with $C(P_0^m)$ given by \eqref{eqn:coP0}. The advantage of the former approach is that it includes a smooth objective. Appendix \ref{sec:appendixa} proves the equivalence by showing that solving \eqref{eqn:mainQP} with $C(P_0^m)$ given by \eqref{eqn:specialcost} ensures that only one of the variables $(P_{0+}^m, P_{0-}^m)$ is nonzero per scenario $m$. This implies that either $P_0^m= P_{0+}^m>0$, or $P_0^m=-P_{0-}^m<0$. For the algorithm that is to be presented, this particular cost function is considered as an example since it is more challenging to deal with.  For a differentiable convex function, only one variable ($P_0^m$) is needed to be considered, which yields simpler updates.  
\end{remark}

\begin{remark}[Optimality of second-order cone stochastic program]
The second-order cone relaxation of OPF has been proved to be exact for tree networks under certain assumptions  \cite{GaLiToLo2015}.  By applying the techniques developed in \cite{GaLiToLo2015}, sufficient conditions under which the inequality in \eqref{eqn:sopofl} holds as equality for the convex stochastic  program \eqref{eqn:mainQP} are derived in Appendix~\ref{sec:appendixb}.
\end{remark}
The two-stage convex  stochastic program \eqref{eqn:mainQP} developed in this section can be solved by centralized algorithms such as interior point methods.  In the next section, a decentralized solver for \eqref{eqn:mainQP} based on ADMM is developed featuring  closed-form updates per node and per scenario. 
\section{Solution Algorithm}
The detailed  design of the decentralized solution algorithm is presented in this section.  An equivalent problem  to \eqref{eqn:mainQP} which is of the general form amenable to application of ADMM is derived in  Subsection \ref{sub:eqpr}.  This equivalent problem includes judiciously designed auxiliary variables that allow  decomposition per node and per scenario.  Subsection \ref{admm}, briefly outlines the ADMM. Finally, in Subsection \ref{sub:updates}, the closed-form updates per node and per scenario are detailed.
\label{sec:algorithm}
\subsection{Equivalent problem}
\label{sub:eqpr}
The only obstacle in fully decomposing \eqref{eqn:mainQP} into separate nodes is the coupling in power flow equations \eqref{eqn:repofl}-\eqref{eqn:vopofl}.  For instance, in \eqref{eqn:repofl} and \eqref{eqn:reapofl}, node $i$ will need to know $P_j^m$, $Q_j^m$, and $l_j^m$ from child nodes $j \in \mathcal{C}_i$.   In order to decouple each node from its child nodes, $N$ respective copies of these variables, $\hat{P}_j^m$, $\hat{Q}_j^m$, and $\hat{l}_j^m$ are introduced per scenario. Moreover, since all nodes except for the root have ancestors, constraint \eqref{eqn:vopofl} is also coupling the nodes.  Therefore, per scenario, another set of $N$ variables $\hat{v}_i^m$ copies $v_{A_i}^m$ at node $i$.    
Finally, an additional set of copies per scenario, namely, $\tilde{P}_i^m$, $\tilde{Q}_i^m$, $\tilde{l}_i^m$, and $\tilde{v}_i^m$ for all $n \in \mathcal{N}\backslash\{0\}$ and the variables $\tilde{P}_{0+}^m$, $\tilde{P}_{0-}^m$ for the root, are also introduced, the purpose of which will be evident shortly. 
\par{} Let the set of boldface variables $\{\mathbf{P},\mathbf{\hat{P}},\mathbf{\tilde{P}}, \mathbf{Q},\mathbf{\hat{Q}},\mathbf{\tilde{Q}}, \mathbf{v},\mathbf{\hat{v}},\\ \mathbf{\tilde{v}},\mathbf{l},\mathbf{\hat{l}},\mathbf{\tilde{l}},\mathbf{p_{c}}, \mathbf{\tilde{p}_c}, \mathbf{q}_{w},\mathbf{\tilde{q}}_{w},\mathbf{P_{0+}},\mathbf{P_{0-}},\mathbf{\tilde{P}_{0+}},\mathbf{\tilde{P}_{0-}}\}$ represent vectors collecting the corresponding variables in all scenarios and nodes. As listed in Table~\ref{table:variables}, these variables are further collected in vectors $\mathbf{x}=\{\mathbf{x}_i^m\}_{i \in \mathcal{N}, m \in \mathcal{M}}$ and $\mathbf{z}=\{\mathbf{z}_i^m\}_{i \in \mathcal{N}, m \in \mathcal{M}}$. 
 
The problem takes the following form:
\begin{subequations}
\label{eqn:equivalent}
\begin{IEEEeqnarray}{rCl}
\min_{\mathbf{x},\mathbf{z}} -\sum\limits_{i=1}^Nu_i(\tilde{p}_{c_i})+\sum\limits_{m=1}^M \pi^m (aP_{0+}^m-bP_{0-}^m)\nonumber \\+
K_\mathrm{Loss}\sum\limits_{m=1}^M\pi^m\sum\limits_{i=1}^Nr_il_i^m 
\end{IEEEeqnarray}
subject to:\\
\textbf{Coupling Constraints} ($i \in \mathcal{N}, m\in\mathcal{M}$):
\begin{IEEEeqnarray}{lllll} 
\label{eqn:copqlv}
i \neq 0: &\quad P_i^m=\tilde{P}_i^m & \quad Q_i^m=\tilde{Q}_i^m & \quad l_i^m=\tilde{l}_i^m & \\
v_i^m=\tilde{v}_i^m & \quad \hat{v}_i^m=\tilde{v}_{A_i}^m & \quad p_{c_i}^m=\tilde{p}_{c_i} & \quad q_{w_i}^m=\tilde{q}_{w_i}^m  \\
\label{eqn:copqlpq}
j \in \mathcal{C}_i:& \quad \hat{P}_j^m=\tilde{P}_j^m & \quad \hat{Q}_j^m=\tilde{Q}_j^m& \quad \hat{l}_j^m=\tilde{l}_j^m\\
\label{eqn:cop0}
 &\quad P_{0+}^m=\tilde{P}_{0+}^m&  \quad P_{0-}^m=\tilde{P}_{0-}^m 
\end{IEEEeqnarray}
\textbf{Individual Equality Constraints} ($i \in \mathcal{N}, m\in\mathcal{M}$): 
\begin{IEEEeqnarray}{lll}
\label{eqn:eqpqpc}
\hspace{-0.4cm}P_i^m&=&\sum\limits_{j \in \mathcal{C}_i}(\hat{P}_j^m+r_j\hat{l}_j^m)+P_{L_i}+p_{c_i}^m-w_i^m \\
\hspace{-0.4cm} Q_i^m&=&\sum\limits_{j \in \mathcal{C}_i}(\hat{Q}_j^m+x_j\hat{l}_j^m)+Q_{L_i}+q_{c_i}^m-q_{w_i}^m-q_{s_i}v_i^m\\
\label{eqn:eqv}
\hspace{-0.4cm}\hat{v}_i^m&=&v_i^m+2(r_iP_i^m+x_iQ_i^m)+(r_i^2+x_i^2)l_i^m ,  i\neq0
\end{IEEEeqnarray}
where $q_{c_i}^m=\left(\sqrt{\frac{1}{\mathrm{PF}_i^2}-1}\right)p_{c_i}^m$.
\begin{IEEEeqnarray}{lll}
\label{eqn:eqp0}
\hspace{-0.4cm}P_0^m&=&P_{0+}^m-P_{0-}^m
\end{IEEEeqnarray}

\textbf{Individual Inequality Constraints} ($i \in \mathcal{N}, m\in\mathcal{M}$)
\begin{IEEEeqnarray}{lll}
&(\tilde{P}_i^m)^2+(\tilde{Q}_i^m)^2 \le (\tilde{v}_i^m)(\tilde{l}_i^m), i \neq0  \label{eqn:ineqpqvl}\\ 
&(1-\epsilon)^2 \le \frac{\tilde{v}_i^m}{v_0} \le  (1+\epsilon)^2, i\neq0 \label{eqn:vtildebounds}\\
&0 \le \tilde{l}_i^m \le l_i^{\max}, i \neq0   \label{eqn:vltilde} \\
\label{eqn:ineqpc}
 &p_{c_i}^{\min} \le \tilde{p}_{c_i} \le p_{c_i}^{\max} \\ 
\label{eqn:ineqq}
&-q_{w_i^m}^{\max} \le \tilde{q}_{w_i}^m \le q_{w_i^m}^{\max} 
\\
\label{eqn:ineqp0}
&\tilde{P}_{0+}^m \ge 0, \tilde{P}_{0-}^m \ge 0. 
\end{IEEEeqnarray}
\end{subequations}

Clearly, \eqref{eqn:equivalent} is equivalent to \eqref{eqn:mainQP}.

\begin{table}[t]
\caption{$\mathbf{x}$ and $\mathbf{z}$ variables for the ADMM algorithm}
\label{table:variables}
\noindent \centering{}
\renewcommand{\arraystretch}{1.5}
\begin{tabular}{|c|c|c|}
\hline 
 & Nodes involved & Variables \tabularnewline
\hline 
$\mathbf{x}_0^m$ & Root  & $\{P_0^m,Q_0^m,P_{0+}^m, P_{0-}^m$\\ 
& & $\{\hat{P}_j^m,\hat{Q}_j^m,\hat{l}_j^m\}_{j \in \mathcal{C}_0}\}$ \tabularnewline
\hline 
$\mathbf{x}_i^m$ & Neither root nor leaf& $\{P_i^m,Q_i^m,v_i^m,l_i^m,$ \\
& & $\{\hat{P}_j^m,\hat{Q}_j^m,\hat{l}_j^m\}_{j \in \mathcal{C}_i},\hat{v}_i^m,p_{c_i}^m,q_{w_i}^m\}$\tabularnewline
\hline 
$\mathbf{x}_{i}^m$ & Leaf & $\{P_i^m,Q_i^m,v_i^m,l_i^m,\hat{v}_i^m,p_{c_i}^m,q_{w_i}^m\}$ \tabularnewline
\hline 
$\mathbf{x}_{i}$ & All nodes & $\{\mathbf{x}_i^m\}_{m=1}^M$ \tabularnewline
\hline
$\mathbf{z}_0^m$ & Root & $\{\tilde{P}_{0+}^m,\tilde{P}_{0-}^m\}$ \tabularnewline 
\hline
$\mathbf{z}_i^m$ & Not root& $\{\tilde{P}_i^m,\tilde{Q}_i^m,\tilde{v}_i^m,\tilde{l}_i^m,\tilde{q}_{w_i}^m\}$ \tabularnewline
\hline
$\mathbf{z}_i$ & Not root & $\{\{\mathbf{z}_i^m\}_{m=1}^M, \tilde{p}_{c_i}\}$ \tabularnewline 
\hline
\end{tabular}
\end{table}
\subsection{Review of ADMM}
\label{admm}
With the previous definitions of $\mathbf{x}$ and $\mathbf{z}$, problem~\eqref{eqn:equivalent} is of the general 
form~\cite{BoPaChPeEc-FnT11}
\begin{align}
\min_{\mathbf{x}\in \mathcal{X},\mathbf{z}\in \mathcal{Z}}~~ & f(\mathbf{x})+g(\mathbf{z}) \quad
\text{subj.~to}~~~~ &  A\mathbf{x}+B\mathbf{z}=\mathbf{c}  \label{admm-coupling} 
\end{align}
where $f$ and $g$ are convex functions.  The set $\mathcal{X}$ corresponds to the individual equality constraints \eqref{eqn:eqpqpc}-\eqref{eqn:eqp0} and $\mathcal{Z}$ captures all the inequality constraints \eqref{eqn:ineqpqvl}-\eqref{eqn:ineqp0}.    

The augmented Lagrangian function is defined as: 
\begin{align}
\label{eqn:augmented}
L_{\rho}(\mathbf{x},\mathbf{z},\mathbf{y})=f(\mathbf{x})+g(\mathbf{z})+\mathbf{y}^T(A\mathbf{x}+B\mathbf{z}-\mathbf{c})\notag \\+\frac{\rho}{2}\|A\mathbf{x}+B\mathbf{z}-\mathbf{c}\|_2^2
\end{align}
where $\mathbf{y}$ is the Lagrange multiplier vector for the linear equality constraints in \eqref{admm-coupling}, and $\rho>0$ is a parameter.   The primal and dual iterations of ADMM are as follows, where $k$ is the iteration index.
\newcommand{\argmin}{\operatornamewithlimits{argmin}} 
\begin{align}
\mathbf{x}{(k+1)} & :=\argmin_{\mathbf{x}\in\mathcal{X}}{L_{\rho}(\mathbf{x},\mathbf{z}(k),\mathbf{y}(k))}\\
\mathbf{z}{(k+1)} & :=\argmin_{\mathbf{z}\in\mathcal{Z}}{L_{\rho}(\mathbf{x}(k+1),\mathbf{z},\mathbf{y}(k))}\\
\mathbf{y}{(k+1)} & :=\mathbf{y}(k)+\rho\left[A\mathbf{x}(k+1)+B\mathbf{z}(k+1)-\mathbf{c}\right]. \label{lagrangeUpdateGeneral}
\end{align}

The purpose of introducing the \textit{tilde} variables in \eqref{eqn:equivalent} is so that the individual inequality constraints in $\mathcal{Z}$ can be handled separately in the $\mathbf{z}$-update.   The $\mathbf{x}$-update on the other hand turns out  to be an equality constrained quadratic program.   This separation of variables are essential to finding closed-form solutions for the updates.
The following primal and dual residuals are measured in every step, and the algorithm is stopped once these are below an acceptable threshold:
\begin{subequations}
\label{eqn:residuals}
\begin{align}
 r(k)&:=||A\mathbf{x}(k)+B\mathbf{z}(k)-c|| \label{eqn:primalres} \\  
 s(k)&:=\rho||A^TB(\mathbf{z}(k)-\mathbf{z}(k-1))|| \label{eqn:dualres}.
\end{align} 
\end{subequations}
\subsection{Updates}
\label{sub:updates}
The Lagrange multipliers corresponding to the coupling constrains of \eqref{eqn:copqlv}-\eqref{eqn:cop0} are listed in Table \ref{table:lagrange}. To perform ADMM, first the Augmented Lagrangian \eqref{eqn:augmented} for problem \eqref{eqn:equivalent} needs to be formed.  This Augmented Lagrangian is separable across variables $\mathbf{x}_i^m$ ($i \in \mathcal{N}$, $m \in \mathcal{M}$) with $\mathbf{z}$ fixed, or across variables $\mathbf{z}_i^m$ and $\tilde{p}_{c_i}$ ($i \in \mathcal{N}$, $m \in \mathcal{M}$) with $\mathbf{x}$ fixed. Each step of the ADMM will consist of minimizing the augmented Lagrangian with respect to either $\mathbf{x}$ or $\mathbf{z}$ and updating the Lagrange multipliers. 
\subsubsection{$\mathbf{x}_i^m$-update}
For node $i \in \mathcal{N} \setminus \{0\}$, per scenario $m$, the $\mathbf{x}$-update will be derived by minimizing the corresponding part of the augmented Lagrangian per node $i$ and scenario $m$ :
\begin{IEEEeqnarray}{lCl}
\label{eqn:xi}
\min_{\mathbf{x}_i^m} \: K_\mathrm{Loss} \pi^m r_il_i^m + \lambda_i^m(P_i^m-\tilde{P}_i^m)+\sum\limits_{j \in \mathcal{C}_i}\hat{\lambda}_j^m(\hat{P}_j^m-\tilde{P}_j^m) \notag \\ +\mu_i^m(Q_i^m-\tilde{Q}_i^m)+ \sum\limits_{j \in \mathcal{C}_i}\hat{\mu}_j^m(\hat{Q}_j^m-\tilde{Q}_j^m) +\gamma_i^m(l_i^m-\tilde{l}_i^m) \notag \\ +\sum\limits_{j \in \mathcal{C}_i}\hat{\gamma}_j^m(\hat{l}_j^m-\tilde{l}_j^m)+\omega_i^m(v_i^m-\tilde{v}_i^m) +\sum\limits_{j\in \mathcal{C}_i}\hat{\omega}_j^m(\hat{v}_j^m-\tilde{v}_i^m) \notag \\+\eta_i^m(p_{c_i}^m-\tilde{p}_{c_i}) +\theta_i^m(q_{w_i}^m-\tilde{q}_{w_i}^m)+\frac{\rho}{2}
\big[(P_i^m-\tilde{P}_i^m)^2  \notag \\
 +\sum\limits_{j \in \mathcal{C}_i}(\hat{P}_j^m-\tilde{P}_j^m)^2  
+(Q_i^m-\tilde{Q}_i^m)^2+\sum\limits_{j \in \mathcal{C}_i}(\hat{Q}_j^m-\tilde{Q}_j^m)^2 \notag 
 \\  + (l_i^m-\tilde{l}_i^m)^2 + \sum\limits_{j \in \mathcal{C}_i}(\hat{l}_j^m-\tilde{l}_j^m)^2 +(v_i^m-\tilde{v}_i^m)^2 \notag  \\  +\sum\limits_{j\in \mathcal{C}_i}(\hat{v}_j^m-\tilde{v}_i^m)^2 +(p_{c_i}^m-\tilde{p}_{c_i})^2+(q_{w_i}-\tilde{q}_{w_i}^m)^2\big] \IEEEeqnarraynumspace
\end{IEEEeqnarray}
subject to \eqref{eqn:eqpqpc} -- \eqref{eqn:eqv}.  

For the special case of $i=0$ , the corresponding Lagrangian will include the term $
\pi^m (aP_{0+}^m-bP_{0-}^m)+ \frac{\rho}{2}[(P_{0+}^m-\tilde{P}_{0+}^m)^2+(P_{0-}^m-\tilde{P}_{0-}^m)^2] $
 with the constraint $P_0^m=P_{0+}^m-P_{0-}^m$.

For all $i \in \mathcal{N}$ problem \eqref{eqn:xi} is of the following form:
\begin{align}
\label{eqn:lineq-opt-closedform}
\frac{1}{2}(\mathbf{x}_{i}^m)^T\mathbf{A}_i^m\mathbf{x}_i^m+(\mathbf{b}_i^m)^T\mathbf{x}_i^m \quad \text{subj. to\ }  \mathbf{C}_i^m\mathbf{x}_i^m=\mathbf{d}_i^m.
\end{align}
The structure of the problem leads to a diagonal $\mathbf{A}_i^m$ and a full-rank $\mathbf{C}_i^m$, and therefore has a closed-form solution:
\begin{align}
\label{eqn:xclose}
\mathbf{x}_i^{m^*}= 
&\mathbf{A}_i^{m^{-1}}(-\mathbf{b}_i^m+\mathbf{C}_i^{m^T}\mathbf{F}_i^m) \\
\text{where} \quad &\mathbf{F}_i^m=(\mathbf{C}_i^m\mathbf{A}_i^{m^{-1}}\mathbf{C}_i^{m^T})^{-1}(\mathbf{d}_i^m+\mathbf{C}_i^m\mathbf{A}_i^{m^{-1}}\mathbf{b}_i^m). \notag
\end{align}
\subsubsection{$\mathbf{z}_i$-update}
In order to find the updates for $\mathbf{z}_i^m$--variables the augmented Lagrangian in \eqref{eqn:augmented} can be minimized separately for  $ \{ \tilde{P}_i^m,\tilde{Q}_i^m,\tilde{v}_i^m,\tilde{l}_i^m \}$, $\tilde{p}_{c_i}$, $\tilde{q}_{w_i}^m$, $\tilde{P}_{0+}^m$, and $\tilde{P}_{0-}^m$ per $i \in \mathcal{N}$ and $m \in \mathcal{M}$ 
subject to \eqref{eqn:ineqpqvl}-\eqref{eqn:ineqp0}.

The minimization with respect to $ \{ \tilde{P}_i^m,\tilde{Q}_i^m,\tilde{v}_i^m,\tilde{l}_i^m \}$ has a closed-form solution which is developed in Appendix~\ref{sec:appendixc} by generalizing \cite[Appendix I]{PeLo14}.

The variable $\tilde{p}_{c_i}$ is not dependent on the scenario, and its update is by solving the following program at every node: 
\begin{align}
\tilde{p}_{c_i}=\argmin_{0 \le \tilde{p}_{c_i} \le p_{c_i}^{\max}} \left[u_i(\tilde{p}_{c_i})+\sum\limits_{m=1}^{M}\eta_i^m(p_{c_i}^m-\tilde{p}_{c_i})\nonumber \right. \\ \left. +\frac{\rho}{2}\sum\limits_{m=1}^{M}(p_{c_i}^m-\tilde{p}_{c_i})^2\right].
\end{align} 
This problem is a scalar box-constrained convex optimization problem, and has a closed-form solution if e.g., the utility function is quadratic.  
The remaining $\mathbf{z}$--variables, namely $\tilde{q}_{w_i}^m$, $\tilde{P}_{0+}^m$, and $\tilde{P}_{0-}^m$, are scenario dependent, and their closed-form updates are given as follows (by minimizing the corresponding term in the Lagrangian):
\begin{IEEEeqnarray}{rCl}
\label{eqn:tildeqwupdate}
\tilde{q}_{w_i}^{m}(k+1)&=&\left[ \frac{\theta_i^m+\rho q_{w_i}^m}{\rho}\right]_{-q_{w_i}^{\max}}^{q_{w_i}^{\max}}
\\
\label{eqn:tildeP0plusupdate}
\tilde{P}_{0+}^m(k+1)&=&\left[\frac{\zeta_+^m+\rho P_{0+}^m}{\rho}\right]^+\\
\label{eqn:tildeP0minusupdate}
\tilde{P}_{0-}^m(k+1)&=&\left[\frac{\zeta_-^m+\rho P_{0-}^m}{\rho}\right]^+
\end{IEEEeqnarray}

where $[t]^+:=\max\{0,t\}$ and $[t]_{t_1}^{t_2}:=\max\left\{t_1,\min\{t,t_2\}\right\}$.

\begin{table}[t]
\caption{Lagrange Multipliers}
\label{table:lagrange}
\noindent \centering{}
\renewcommand{\arraystretch}{1.5}
\begin{tabular}{|c|c|}
\hline 
 Equality Constraint & Lagrange Multiplier \tabularnewline
\hline 
$P_i^m=\tilde{P}_i^m$ & $\lambda_i^m$ \tabularnewline
\hline
$Q_i^m=\tilde{Q}_i^m$ & $\mu_i^m$  \tabularnewline
\hline
$l_i^m=\tilde{l}_i^m$ & $\gamma_i^m$ \tabularnewline 
\hline
$v_i^m=\tilde{v}_i^m$ & $\omega_i^m$ \tabularnewline
\hline
$\hat{v}_j^m=\tilde{v}_{A_j}^m$ & $\hat{\omega}_j^m$ \tabularnewline
\hline
$\hat{P}_j^m=\tilde{P}_{j \in \mathcal{C}_i}^m$ & $\hat{\lambda}_j^m$ \tabularnewline
\hline
$\hat{Q}_j^m=\tilde{Q}_{j \in \mathcal{C}_i}^m$ & $\hat{\mu}_j^m$ \tabularnewline
\hline
$\hat{l}_j^m=\tilde{l}_{j \in \mathcal{C}_i}^m$ & $\hat{\gamma}_j^m$ \tabularnewline
\hline
$p_{c_i}^m=\tilde{p}_{c_i}$ & $\eta_i^m$  \tabularnewline
\hline
$q_{w_i}^m=\tilde{q}_{w_i}^m$ & $\theta_i^m$ \tabularnewline 
\hline
$P_{0+}^m=\tilde{P}_{0+}^m$ & $\zeta_{+}^m$ \tabularnewline
\hline
$P_{0-}^m=\tilde{P}_{0-}^m$ & $\zeta_{-}^m$ \tabularnewline
\hline
\end{tabular}
\end{table}

Following the detailed derivation of the ADMM-based algorithm in the present section, the next section deals with the   implementation of the algorithm in a distributed fashion, and highlights its advantages. 
\section{Decentralized Implementation}
\label{sec:commreq}
In the implementation of this algorithm, each node $i \in \mathcal{N}$ is  responsible for maintaining and updating variables $\mathbf{x}_i$, $\mathbf{z}_i$, and the corresponding Lagrange multipliers.

\begin{figure}[t]
  \centering{
  \vspace{0.2cm}
      \includegraphics[scale=0.35]{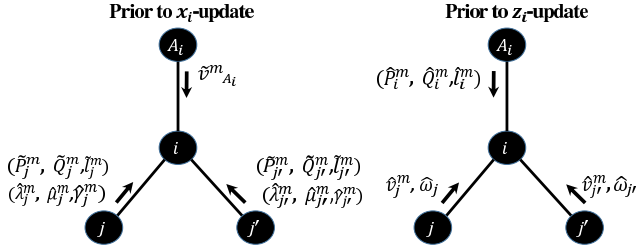}}
  \caption{Communication requirements of the ADMM algorithm. In each iteration, prior to the $\mathbf{x}$-update, node $i$ receives $\tilde{P}_j^m$, $\tilde{Q}_j^m$, and $\tilde{l}_j^m$ from its children nodes,   and receives $\tilde{v}_{A_i}^m$ from its ancestor node.  Prior to the $\mathbf{z}$-update, ancestor node $A_i$ sends $\hat{P}_i^m$, $\hat{Q}_i^m$, and $\hat{l}_i^m$ to node $i$ while node $i$ receives all the  $\hat{v}_{j}^m$ variables from its children. Note that whenever a variable is transmitted, its corresponding Lagrange multiplier is transmitted as well.}   \label{fig:commReqTree}
\end{figure}

The ADMM algorithm works as depicted in Fig.~\ref{fig:commReqTree}.  First, $\mathbf{z}$ and the Lagrange multipliers are initialized with arbitrary numbers. In each iteration, every node $i$ that has children receives $\tilde{P}_j^m$, $\tilde{Q}_j^m$, and $\tilde{l}_j^m$ from all $j \in \mathcal{C}_i$. Also, each node $i$ receives $\tilde{v}_{A_i}^m$ from its ancestor.  Using these variables, $\mathbf{x}_i^m$  is updated according to the closed-form solution in \eqref{eqn:xclose}. 

Prior to the $\mathbf{z}$-update step, $\hat{P}_i^m, \hat{Q}_i^m$, and $\hat{l}_i^m$ are sent to node $i$ from ancestor $A_i$, and node $i$ collects $\{\hat{v}_j^m\}_{j \in \mathcal{C}_i}$ from its children.  Upon receiving the required information, node $i$ performs the $\mathbf{z}$-update step.  Upon completion of the $\mathbf{z}$-update step, the Lagrange multipliers are updated.   

Note that, node 0 only communicates with its children, and the leaf nodes only communicate with their ancestors.  All other nodes communicate both with their children and ancestors.  Therefore in this algorithm only neighbors will need to communicate.   

Algorithm~\ref{alg:1} summarizes the algorithm.  Convergence in Step 2 of the algorithm is declared when the residuals $r(k)$, $s(k)$  in \eqref{eqn:residuals} are sufficiently small and the value $\max_{i,m}\{v_i^ml_i^m-(P_i^m)^2-(Q_i^m)^2\}$ is smaller than $10^{-3}$ $(\mathrm{pu})^2$.  The latter ensures exactness of the SOCP relaxation in \eqref{eqn:sopofl}.

This section is wrapped up by highlighting the  merits of the developed algorithm:
\begin{enumerate}
\item The algorithm comprises closed-form updates per node and per scenario. The need for solving complex optimization problems per node is therefore bypassed which greatly simplifies implementation.
\item Computational effort per node does not change as the size of the
network increases---that is, each node still needs to run the same closed-form updates.
\item This distributed algorithm is conducive to maintaining user privacy.  Specifically, global optimality is achieved while parameters such as bounds on user consumption (i.e., $p_{c_i}^{\min},  p_{c_i}^{\max}$ ), power factor, and user utility are not transmitted to a central agent.
\end{enumerate}

\begin{algorithm}
\caption{Required Communications and Updates}
\label{alg:1}
\begin{algorithmic}[1]
\State Initialize $\mathbf{z}$-variables and Lagrange multipliers with random numbers at every node $i$.
\State For every node $i$ repeat steps \ref{step1}-\ref{stepEnd}  until convergence. 
\label{step1} 
\State Receive $\tilde{P}_j^m$, $\tilde{Q}_j^m$, $\tilde{l}_j^m$, $\hat{\lambda}_j^m$, $\hat{\mu}_j^m$, and $\hat{\gamma}_j^m$ from all $j \in \mathcal{C}_i$ and for $m \in \mathcal{M}$.   Also receive $\tilde{v}_{A_i}^m$ from node $A_i$ and $m \in \mathcal{M}$.
\State Perform $\mathbf{x}_i$-update.
\State Receive the updated $\mathbf{x}$-variables $\hat{P}_i^m$, $\hat{Q}_i^m$ and $\hat{l}_i^m$ for  $m \in \mathcal{M}$ from $A_i$. Also receive $\hat{v}_j^m$ and $\hat{\omega}_j^m$ from all nodes $j \in \mathcal{C}_i$ and $m \in \mathcal{M}$.
\State Perform $\mathbf{z}_i$-update. 
\State Update the Lagrange multipliers. \label{stepEnd} 
\end{algorithmic}
\end{algorithm}

\section{Numerical Tests}
\label{sec:numtest}
\subsection{Network setup}
\label{sec:numtesta}
\begin{figure}[t]
\centering
\includegraphics[scale=0.3]{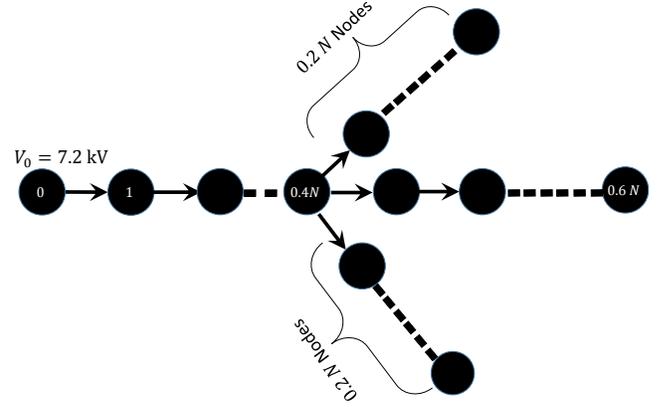}
\caption{Radial distribution network used in the numerical tests.}
\label{fig:twolateral}
\end{figure}

Numerical simulations are conducted on a sample tree distribution network as illustrated in Fig.~\ref{fig:twolateral}.  Resistance and reactance values on line $i$, i.e., $r_i+jx_i$,  are considered constant and equal to $0.33+j0.38 \frac{\Omega}{\mathrm{km}} \times d (\mathrm{km})$ where $d$ represents the distance between two nodes (fixed at $0.2 \mathrm{km}$).   For the network, $S_{\mathrm{base}}=1$ MVA is selected, while the substation voltage is fixed at $V_0=7.2$ kV.  Voltage regulation is performed with $\epsilon=0.05$ so that nodal voltages are allowed to vary within $5 \%$ of the nominal value $V_0$. The line flow limit is $l_i^{\max}=0.5  (\mathrm{kA})^2$. 
The cost $C(P_0^m)$ is set to zero, and similar to \cite{LiChLo2012}, the utility function is selected as  $u_i(p_{c_i})=-K_{u_i}(p_{c_i}-p_{c_i}^{\max})^2 $.The objective weights are $K_{u_i}=1$ and $K_{\mathrm{Loss}}=1$.  Notice that if we set $K_{\mathrm{Loss}}=0$, then $C(P_0^m)$ can correctly account for loss terms.

For users $i=1,\ldots,N$, the non-elastic load is $P_{L_i}=0.1 \ \mathrm{MW}$, while $p_{c_i}$  is constrained to be in $[0,p_{c_i}^{\max}]=[0,0.05] \ \mathrm{MW}$.  The  power factor for both elastic and non-elastic loads is selected to be $\mathrm{PF}_i=0.94$.
 
 \subsection{Methodology for generating scenarios}
 \label{sec:numtestb}
If there is a PV unit at node $i$, then the relationship between $s_{w_i}$ and the maximum real power capability of the inverter $w_i^{\max}$  is given by $w_i^{\max}=\frac{s_{w_i}}{1.1}$.   Following \cite{WaChWaBe2014} and \cite{NiZaAg2012},  the actual power generated by the PV unit, $w_i$, is a random variable that takes values from a beta distribution  with mean $\bar{w}_i$ and variance $\sigma_i^2$, as follows:
\begin{subequations}
\label{eqn:beta}
\begin{IEEEeqnarray}{rCl}
&f_{W_i}(w_i)=\frac{1}{w_i^{\max}}(\frac{w_i}{w_i^{\max}})^{\alpha-1}(1-\frac{w_i}{w_i^{\max}})^{\beta-1} \\
&\mathrm{Mean}=\bar{w}_i=\frac{\alpha}{\alpha+\beta}w_i^{\max}  \label{eqn:betaMean}\\
&\mathrm{Variance}=\sigma_i^2=\frac{\alpha \beta}{(\alpha+\beta)^2.(\alpha+\beta+1)}(w_i^{\max})^2  \label{eqn:betavar}
\end{IEEEeqnarray}
where the relationship between the mean and the standard deviation is given as: 
\begin{IEEEeqnarray}{rCl}
\frac{\sigma_i}{w_i^{\max}}=0.2\frac{\bar{w}_i}{w_i^{\max}}+0.21.
\end{IEEEeqnarray}
\end{subequations}
In each of the ensuing case studies, it is assumed that $\bar{w}_i$ is known, and subsequently $\alpha$ and $\beta$ can be found numerically using \eqref{eqn:betaMean} and \eqref{eqn:betavar}. Then, 1000 equiprobable scenarios are generated according to \eqref{eqn:beta}. This number is then reduced to $M=7$ representative scenarios using the fast forward reduction method \cite{Conejo-Uncert}, \cite{WaChWaBe2014}. 

The scenario reduction methodology starts with an original scenario set $\Omega$ with cardinality $|\Omega|=1000$ and an empty set $\Omega_s=\emptyset$. Then, in each iteration, a scenario from $\Omega \backslash \Omega_s$  is selected which minimizes the Kantorovich Distance between $\Omega$ and $\Omega_s$ \cite{Conejo-Uncert}.  The algorithm stops once the prescribed cardinality $|\Omega_s|=M=7$ is achieved. The probability of scenarios that are not included in the reduced representative set (i.e., $\Omega_s$) are aggregated on to the probability of the closest representative scenario in $\Omega_s$.  The choice for $M=7$ is so that the smallest probability in the reduced set $\Omega_s$ is at least $0.01$.  

\begin{remark}[Number of scenarios in networks with $N$ distributed generators] In a network with $N$ generators whose power injections come from independent distributions, an exponential number of scenarios would potentially be needed to formulate the stochastic program. However, the stochastic program in this paper will not suffer from this exponential growth because of the following reasons:  1) Generators are all in a geographically limited area and under similar irradiance conditions, and hence generator outputs are spatially correlated; 2) the forecasted power output for the next time period---ranging from e.g., 15 minutes to 1 hour---is available, and is used to set the mean value of the distribution for each injection.  The resulting distributions will have similar parameters, according to the type of day (e.g., sunny, cloudy, partly cloudy). \end{remark} 

Different scenarios are generated for each of the case studies that follow, and problem \eqref{eqn:equivalent} is solved.  It is numerically verified for all problems that the SOCP inequality of \eqref{eqn:sopofl} holds as equality for every scenario.

\subsection{Case study for different day types}
\label{subsec:numc}
In this case study, the network features $N=50$ nodes with 30 nodes on the main branch and two laterals of 10 nodes each branching off at node 20. Nodes 31 to 40 correspond to the first branch while nodes 41 to 50 correspond to the second branch. All users are equipped with distributed PV generators and one large PV generation unit is located at the terminal node of the main branch.

For the larger PV installation in the terminal node of the main branch, $s_{w_{30}}=1$ MVA is selected, while for the remaining nodes we set $s_{w_i}=0.1$ MVA. The ratio of $\frac{\bar{w}_i}{w_i^{\max}}$ takes the values $0.3, 0.6$, and $0.9$, corresponding respectively to cloudy, partly cloudy and sunny days.
\begin{table}[t]\footnotesize
\caption{Objective Value Breakdown for the Three Day Types}
\centering
\renewcommand{\arraystretch}{2}
\begin{tabular}{|p{3.3cm}|c|c|c|} 
\hline
\diagbox{Objective}{Day Type} & Cloudy  &  Partly Cloudy & Sunny  \\
\hline \hline
\begin{tabular}{c} Negative Utility \\ $-\sum\limits_{i=1}^Nu_i(p_{c_i})$ monet. units \end{tabular}  &   0.1142 &    0.0689 & 0.0232 \\
\hline
\begin{tabular}{c} Expected Thermal Losses \\ $\sum\limits_{m=1}^M \pi^m \sum\limits_{i=1}^N r_il_i^m$ ($\mathrm{MW}$) \end{tabular} & 0.3181 &  0.0664 &    0.0271\\
\hline
\begin{tabular}{c} Objective Value Total \\ ($\times 10^6$) \end{tabular} & 0.4323 &  0.1357 &    0.0503\\
\hline
\end{tabular}
\label{table:daytypeComparisons}
\end{table}
\begin{figure}[t]
  \centering{
      \includegraphics[scale=0.41]{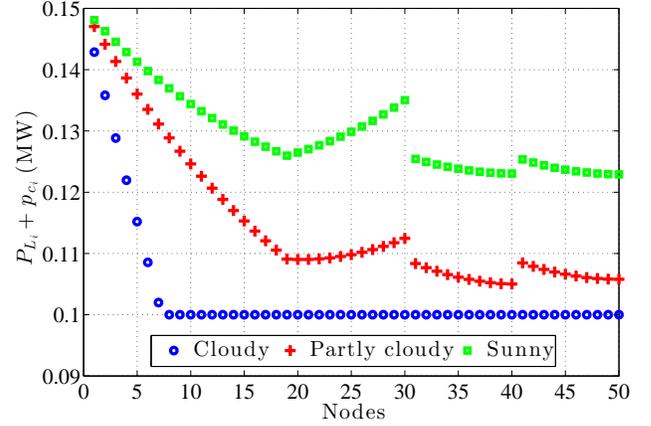}} 
     \caption{Optimal user consumption values $P_{L_i}+p_{c_i}$ in three different day types.  Level of satisfied demand increases as conditions range from cloudy to sunny.}
\label{fig:pcday}
\end{figure}

Table~\ref{table:daytypeComparisons} shows the breakdown of objective values for problem~\eqref{eqn:equivalent} solved for the three different day types.  As conditions range from cloudy to sunny, it is observed that the level of satisfied elastic demand increases, while the expected thermal losses decrease.  In particular, Fig.~\ref{fig:pcday} shows the total load scheduled for each user in these three day types.  On cloudy days, i.e., when PV generation is low, the proposed stochastic program ensures meeting the non-elastic demand.  As PV generation increases during partly cloudy and sunny days, larger portions of the elastic demand are also guaranteed.

\begin{figure}[t]
  \centering{
      \includegraphics[scale=0.41]{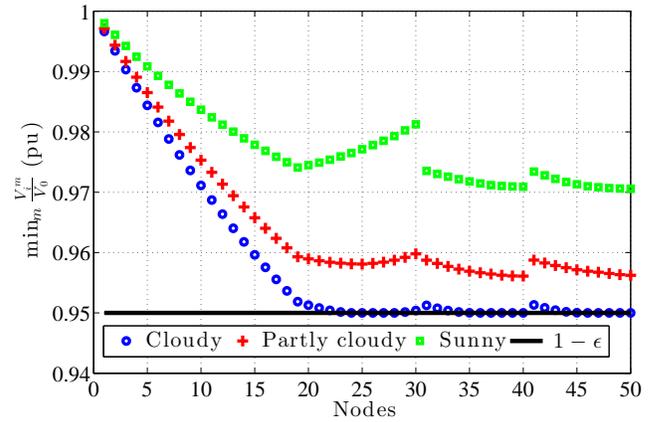}} 
\caption{Worst-case voltage profile across all scenarios for the three different day types.  In a cloudy day, the voltage drop is high.  As PV generation rises in partly cloudy and sunny days, voltage drops decrease.}
\label{fig:averageabs}
\end{figure}

Fig.~\ref{fig:averageabs} depicts the worst-case voltage profile across all scenarios for the three different day types. There are three break points for each voltage plot.   The voltage rise at node 30 corresponds to the larger generation of the terminal node.  Voltage drop at node 20 corresponds to the branching of the network.  On a cloudy day, the reduced power generation in the network results in a higher voltage drop.  This voltage drop is smaller for improved solar conditions.

\subsection{Comparison with distributed local control of \cite{TSBC10b}}
The second set of numerical tests are conducted on a larger network with $N=100$ nodes comprising $60$ nodes on the main branch and two laterals each of length 20 branching off at node $40$.   In this setup, $50 \%$ of the user nodes (randomly selected) are capable of PV generation. The terminal node on the main branch also provides large injection with $s_{w_{60}}=1$ MVA, while $s_{w_i}=0.4$ MVA is selected for  the smaller generations.  The simulations are performed for a sunny day. All other parameters are as listed in Subsections \ref{sec:numtesta} and \ref{sec:numtestb}.

In this case, problem~\eqref{eqn:equivalent} is solved first with $N=100$, $M=7$ scenarios and the optimal demand response schedules (i.e., $p_{c_i}$'s) are obtained. Then, in the online phase,  problem \eqref{eqn:equivalent} with $p_{c_i}$'s fixed to the previously found values is solved for a newly generated set of 100 scenarios.  The results are compared to the ones obtained by the local reactive power control policy proposed in \cite{TSBC10b}.
In this scheme, only the local variables $w_i^m$, $p_{c_i}$, and $q_{c_i}$, are used to set $q_{w_i}^m=F_i(w_i^m,P_{c_i},Q_{c_i})$, where 
\begin{IEEEeqnarray}{lll}
F_i(w_i^m,P_{c_i},Q_{c_i})=
\left[\mathrm{K}F_i^{(L)}+(1-\mathrm{K})F_i^{(V)}\right]_{-q_{w_i}^{\max}}^{q_{w_i}^{\max}} \label{locContrK}\\ 
F_i^{(L)}=\left[Q_{c_i}\right]_{-q_{w_i}^{\max}}^{q_{w_i}^{\max}}\\
F_i^{(V)}=\left[Q_{c_i}+\frac{x_i(P_{c_i}-w_i^m)}{r_i}\right]_{-q_{w_i}^{\max}}^{q_{w_i}^{\max}}
\end{IEEEeqnarray}
with $P_{c_i}=P_{L_i}+p_{c_i}$ and $Q_{c_i}=Q_{L_i}+q_{c_i}$. 

This local control policy considers $p_{c_i}$ and $q_{c_i}$ to be specified and does not optimize the user consumption.  Therefore, the optimal $p_{c_i}$ and $q_{c_i}$ values previously obtained by the solution of problem~\eqref{eqn:equivalent} are set as inputs to the local algorithm~\eqref{locContrK}.  Finally, upon setting $q_{w_i}^m$, the power flows $P_i^m$ and $Q_i^m$ as well as the voltages $V_i^m$ can be found through solving the nonlinear power flow equations using Newton's method \cite{loadflow}. Notice that the parameter $K \in  \mathbb{R}$ in \eqref{locContrK}  also needs to be experimentally set via trial and error.

Table \ref{table:compareVsTuritsyn} lists the thermal losses and the maximum voltage deviation resulting from the stochastic programming approach and the local control policy for different values of $K$.
\begin{table}[t]
\caption{Objective Values, Max Voltage Deviation and Average Voltage Deviation}
\centering
\renewcommand{\arraystretch}{2}
\begin{tabular}{|c|c|c|}
\hline
Method &  Loss (MW)  & $\max_{i,m}\frac{|V_i^m-V0|}{V0}$ (pu)\\
\hline \hline
Stoch. Progr. &   0.0940  &   0.0500\\
\hline
$K=1.1$ &    0.1927 &   0.3682 \\
\hline
$K=1.2$ &    0.1229 &   0.3098 \\
\hline
$K=1.3$ & 0.1005 & 0.2647 \\
\hline
$K=1.4$ &   0.1155 &  0.2266 \\
\hline
$K=1.5$ & 0.1616 &  0.1931 \\
\hline
$K=1.6$ &  0.2341 & 0.2018 \\
\hline
$K=1.7$ & 0.2674 & 0.2154 \\
\hline
$K=1.8$ & 0.2622 & 0.2166 \\
\hline
$K=1.9$ & 0.2539 & 0.2166 \\
\hline
$K=3$ & 0.2440 & 0.2160 \\
\hline
\end{tabular}
\label{table:compareVsTuritsyn}
\end{table}
\begin{figure}[t]
\centering
\includegraphics[scale=0.35]{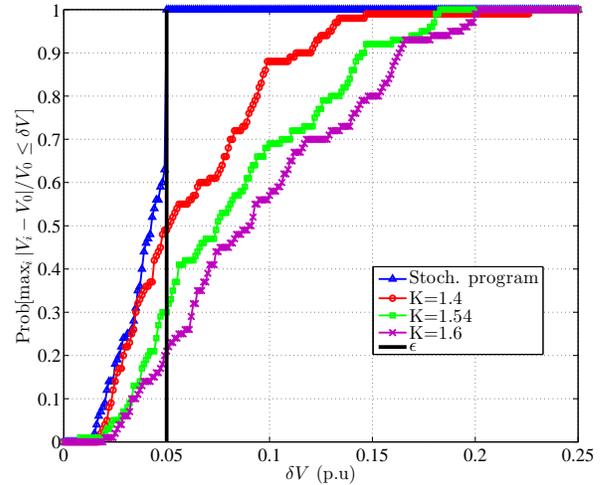}
\caption{Empirical cumulative distribution function of the maximum voltage deviation, i.e.,  $\max_i \frac{|V_i-V_0|}{V_0}$, for the proposed stochastic model as well as local control policy with several values of $K$.  Ideally, it is preferred to have the CDF plots to be on the left side of the solid $\epsilon$ line which corresponds to voltage deviations below $\epsilon$.  }
\label{fig:comp}

\end{figure}

The table reveals that for certain values of $K$ (such as $K=1.3$) the local control policy performs well in terms of thermal losses---partially due to the fact that the inputs to~\eqref{locContrK} are the optimal real power consumptions.  However, the local control policy fails to guarantee that voltage levels are within the $\epsilon$ range.   The best $K$ in terms of voltage regulation was found with a grid search to be $K=1.54$, resulting in a voltage deviation of 0.18 pu, which violates the voltage constraint, and in thermal losses of 0.18 MW---which is two times greater than that of the proposed stochastic programming approach.

To get a closer look at the voltage regulation, the empirical cumulative distribution function (CDF) of the maximum voltage deviation across nodes, i.e.,  $\max_{i}\frac{|V_i-V_0|}{V_0}$, is plotted in Fig.~\ref{fig:comp} for the stochastic programming approach as well as for different values of $K$.  The CDF is obtained by counting the number of scenarios in the online phase for which $\max_i \frac{|V_i-V_0|}{V_0}$ is less than $\delta V$ and dividing this number by the total number of 100 test scenarios.  It is seen that the stochastic program, even by only considering just a few scenarios,  guarantees the maximum voltage deviation to be less than the required threshold in all instances whereas the CDF induced by the local control policy reveals that voltage deviations may exceed the required threshold.
\begin{table}[t]
\caption{Number of infeasible scenarios when reactive power compensation of PV inverters is not allowed}
\centering
\renewcommand{\arraystretch}{2}
\begin{tabular}{|c|c|c|c|c|c|}
\hline
PV Penetration (\%) & 30 & 55 & 60 & 65 & 70  \\
\hline
No. Infeasible cases & 100 & 92 & 61 & 29 & 8 \\
\hline
\end{tabular}
\label{table:varcontrol}
\end{table}

\subsection{Case study on a feeder with shunt capacitors}
To show effectiveness of inverter reactive power control, the 56-node network of \cite[Fig.~2]{FNCL12} that already includes shunt capacitors is selected for additional numerical tests.  The details of the network are given in \cite[Table I]{FNCL12}.  
Only non-elastic load is considered in this case study (i.e., $p_{c_i}=0$).  The values for $P_{L_i}$'s and $Q_{L_i}$'s are calculated using the apparent peak load in \cite[Table I]{FNCL12} increased by 50 \% because the original network is lightly loaded, and assuming $\mathrm{PF}_i=0.94$ per node.  
 All nodes are PV-enabled.  By defining PV penetration level as the ratio of  total PV apparent power capability to the total load,  $s_{w_i}$'s per node are varied so that different PV penetration levels can be simulated.

 Problem \eqref{eqn:mainQP} is first solved for 100 PV generation scenarios when reactive power compensation by PV inverters is not allowed (i.e., $q_{w_i}^m=0$).   The number of infeasible scenarios is recorded in Table~\ref{table:varcontrol}.  For the same scenarios, problem \eqref{eqn:mainQP} is solved with reactive power compensation capability as in \eqref{eqn:reabo}.  In this case, all scenarios were feasible. 

\begin{figure}[t]
\centering
\includegraphics[scale=0.45]{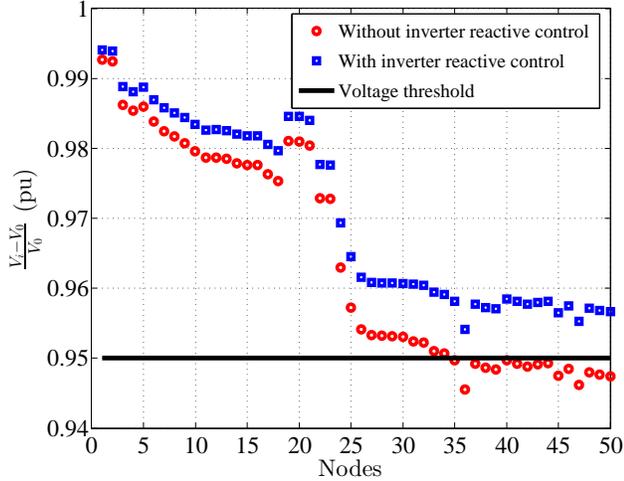}
\caption{Voltage profile   $\frac{|V_i-V_0|}{V_0}$ for a fixed scenario. When inverters do not provide reactive power compensation, the voltage drop may exceed the threshold; while allowing reactive power compensation can prevent this drop.}
\label{fig:varControlcomp}
\end{figure}

When inverters are not allowed to compensate for reactive power, voltages may drop below the required threshold, while reactive power provided by the inverters prevents large voltage drops and increases system reliability.  The voltage profile for a scenario with $w_i=\bar{w}_i$ and $\sum_i\bar{w}_i=0.5\sum_i P_{L_i}$ (i.e., 50 \% penetration) is also plotted in Fig.~\ref{fig:varControlcomp} to illustrate this effect.

\subsection{Effect of stepsize $\rho$ in convergence of ADMM}
This subsection numerically investigates the effect of the stepsize $\rho$  on the convergence of the ADMM algorithm.  In particular, the network setup in Subsection \ref{sec:numtesta} is considered here, where $50 \%$ of nodes are capable of PV generation with $s_{w_i}=0.15$ MVA, and $s_{w_{50}}=1.5$ MVA for the terminal node at one of the laterals. Moreover, $\frac{\bar{w}_i}{w_i^{\max}}=0.75$ is selected.  The remaining parameters are selected as described in Subsections \ref{sec:numtesta} and \ref{sec:numtestb}. Initially, 1000 scenarios are generated, which are subsequently reduced to $M=7$ scenarios.  Problem \eqref{eqn:equivalent} is then solved using the ADMM algorithm of Section~\ref{sec:algorithm} for three constant stepsizes, namely $\rho=2, 20, 100$, and one adaptive stepsize ($\rho_{\mathrm{adaptive}}$)  according to the following rule \cite{BoPaChPeEc-FnT11}: 
\begin{IEEEeqnarray}{rCl}
\rho (k+1):= \begin{cases} 2 \rho(k)  &\text{if} ||r(k)|| > 10 ||s(k)||\\
									 \frac{\rho(k)}{2}  &\text{if} ||s(k)|| > 10 ||r(k)|| \\ 
									 \rho(k) &\text{otherwise} \end{cases}
\end{IEEEeqnarray}
\noindent where $\rho(1)=100$, and the primal and dual residuals, i.e., $r(k)$ and  $s(k)$, are respectively calculated via \eqref{eqn:primalres} and \eqref{eqn:dualres}. 
\begin{figure}[t]
\centering
\includegraphics[scale=0.45]{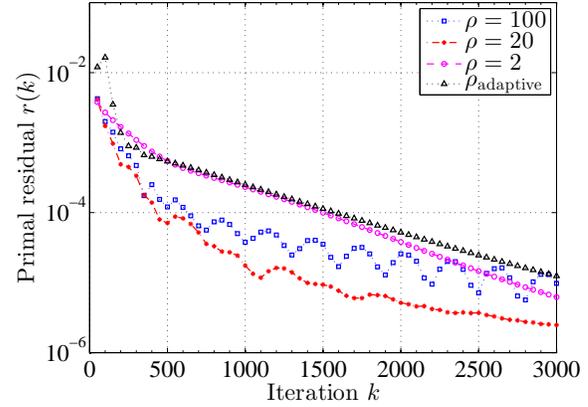}
\caption{Primal residual per ADMM iteration for various stepsizes  ($\rho=2,20,100$ and $\rho_{\mathrm{adaptive}}$).}
\label{fig:diffrhoRerr}
\end{figure}
\begin{figure}[t]
\centering
\includegraphics[scale=0.45]{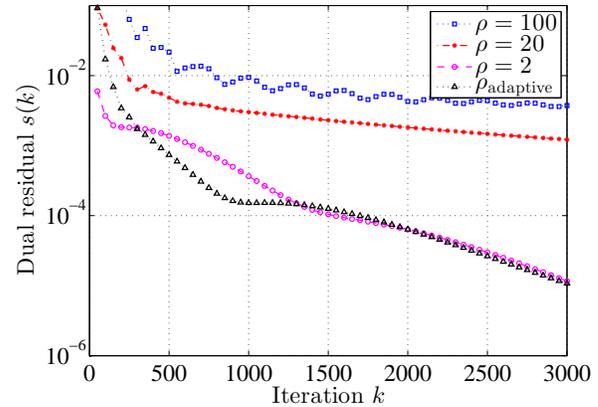}
\caption{Dual residual per ADMM iteration for various stepsizes  ($\rho=2,20,100$ and $\rho_{\mathrm{adaptive}}$).}
\label{fig:diffrhoSerr}
\end{figure}

The resulting primal and dual residuals per iteration are respectively given in Fig.~\ref{fig:diffrhoRerr} and \ref{fig:diffrhoSerr} for the various values of $\rho$.  All choices of $\rho$ (including the adaptive one) perform well in terms of reducing the primal residual, while $\rho=20$ and $\rho=100$ perform poorly in minimizing the dual residual [potentially due to the multiplication in \eqref{eqn:dualres}].  Moreover, $\rho=2$ and $\rho_{\mathrm{adaptive}}$ have a similar performance, noting that the $\rho_{\mathrm{adaptive}}$ reaches 1.56 upon convergence. 

\begin{figure}[t]
\centering
\includegraphics[scale=0.45]{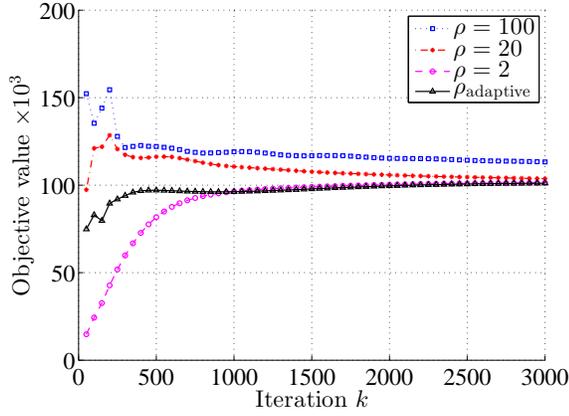}
\caption{Comparison of objective value per ADMM iteration for various stepsizes ($\rho=2,20,100$ and $\rho_{\mathrm{adaptive}}$).}
\label{fig:diffrhoObj}
\end{figure}

\begin{figure}[t]
\centering
\includegraphics[scale=0.45]{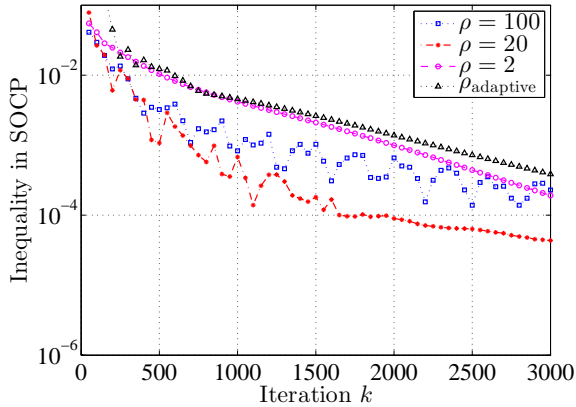}
\caption{Exactness of the SOCP relaxation, i.e., $\max_{i,m} |(P_i^m)^2+(Q_i^m)^2-v_i^ml_i^m|$, per ADMM iteration for various stepsizes ($\rho=2,20,100$ and $\rho_{\mathrm{adaptive}}$).}
\label{fig:diffrhoSOCP}
\end{figure}

The objective value per iteration is shown in Fig.~\ref{fig:diffrhoObj} for the various values of $\rho$.  For $\rho=100$,  an accurate objective value is not found within the 3000 iterations.  The exactness of the SOCP relaxation, i.e., $\max_{i,m} |(P_i^m)^2+(Q_i^m)^2-v_i^ml_i^m|$, is depicted in Fig.~\ref{fig:diffrhoSOCP}.  This value eventually approaches zero for all values of $\rho$; however, the progress is rather slow after 1000 iterations in the case of $\rho=2$ or $\rho_{\mathrm{adaptive}}$. 


\subsection{Effect of number of scenarios on convergence}
This section investigates the convergence of ADMM under larger number of scenarios. Problem \eqref{eqn:equivalent} is solved for the network of Fig.~\ref{fig:twolateral}.  The  penetration level is set to $50 \%$, while $s_{w_i}=0.15$ (MVA) is selected for PV-enabled nodes, and $s_{w_{50}}=1.5$  (MVA) is chosen for the terminal node at one lateral. Problem \eqref{eqn:equivalent} is solved using  $M=100$ and $M=500$ randomly generated  equiprobable scenarios with $\frac{\bar{w}_i}{w_i^{\max}}=0.75$.  Table~\ref{table:scalability} lists the parameters indicating the convergence in these two test cases.  The algorithm scalability is not necessarily dependent on the number of scenarios, but rather on the network structure and the specific power injections per node.  As long as each node is capable of performing individual updates for the specified number of scenarios, the algorithm will converge. 
\begin{table}
\centering 
\caption{Convergence of ADMM for Increased  Number of Scenarios}
\renewcommand{\arraystretch}{1.3}
\begin{tabular}{|c|c|c|}
\hline
M & 100 & 500\\
\hline
\hline
Number of iterations  & 3059 & 4100\\
\hline
Primal residual $r(k)$ & $9.9 \times 10^{-6}$  & $2.23 \times 10^{-5}$ \\
\hline
Dual residual $s(k)$ & $1.24 \times 10^{-6}$ &    $4.4 \times 10^{-5}$\\
\hline
$\max_{i,m} |(P_i^m)^2+(Q_i^m)^2-v_i^ml_i^m|$ & $8.24 \times 10^{-4}$ & 
 $9.38 \times 10^{-4}$ \\
\hline
\end{tabular}
\label{table:scalability}
 \end{table}

\section{Conclusion}
\label{sec:conc}
This paper developed a stochastic power management framework for radial distribution networks with high levels of PV penetration.   Decision variables included real power consumption of programmable loads in user nodes and the reactive power generation or consumption of the PV inverters.  The uncertain real power injections of the user buses were modeled as random variables taking values from a finite number of scenarios.   A convex stochastic optimization program was formulated to minimize the sum of negative utility,  the expected value of cost of power provision, and the expected thermal losses subject to the SOCP relaxation of the power flow equations, power consumption constraints, and voltage regulation specifications.  A decentralized method using the ADMM was developed to solve the stochastic program, in which the updates per node and per scenario turn out to be in closed form.

\appendices

\section{Proof that at most one of the two variables $P_{0+}^m$ and $P_{0-}^m$ is nonzero}
\label{sec:appendixa}

Let $\tilde{P}_{0+}^m$ and $\tilde{P}_{0-}^m$ be the solution of \eqref{eqn:mainQP} with $C(P_0^m)$ replaced by $aP_{0+}^m-bP_{0-}^m$.  Suppose that  $\tilde{P}_{0+}^m > 0$ and $\tilde{P}_{0-}^m > 0$.  Then, $\tilde{P}_{0+}^m-\epsilon$ and $\tilde{P}_{0-}^m- \epsilon$ are feasible for sufficiently small $\epsilon > 0$ and give an objective $a\tilde{P}_{0+}^m-b\tilde{P}_{0-}^m-(a-b)\epsilon$ which is strictly smaller than $a\tilde{P}_{0+}^m-b\tilde{P}_{0-}^m$ since $a > b$.   This is a contradiction.

\section{On the exactness of the SOCP Relaxation}
\label{sec:appendixb}
Based on \cite{GaLiToLo2015}, this appendix presents conditions under which the optimal solution to problem \eqref{eqn:mainQP} satisfies \eqref{eqn:sopofl} with equality.  In order to state the results, some notations are introduced next.  

Given net nodal consumptions $(P_{L_i}+p_{c_i}-w_i^m, Q_{L_i}+q_{c_i}-q_{w_i}^m)$,  the solution to the \texttt{LinDistFlow} approximation of the power flow equations presented in \eqref{eqn:repofl}--\eqref{eqn:vopofl} for  scenario $m$ is given by: 
\begin{IEEEeqnarray}{rCl}
\breve{P}_i^m(\mathbf{p}_{c}) &=&\sum\limits_{j: i \in \mathcal{P}_j} (P_{L_j}
+p_{c_j}-w_j^m) \label{eqn:phatpcipl} \\
\breve{Q}_i^m(\mathbf{p}_{c}, \mathbf{q}_{w}^m) &=& \sum\limits_{j: i \in \mathcal{P}_j}(Q_{L_j}+q_{c_j}-q_{w_j}^m)\label{eqn:qhatqciql}\\
\breve{v}_i^m(\mathbf{p}_{c},\mathbf{q}_{w}^m)  &=& v_0 -2 \sum\limits_{j \in \mathcal{P}_i^+} [r_j\breve{P}_j^m+x_j\breve{Q}_j^m]
\end{IEEEeqnarray}
where $q_{c_j}=\left(\sqrt{\frac{1}{\mathrm{PF}_j^2}-1}\right) p_{c_j}$, $\mathcal{P}_j$ is the unique path from the root node to node $j$ (including node $j$), and $\mathcal{P}_j^+$ is $\mathcal{P}_j \backslash \{0\}$. Also, $\mathbf{p}_{c}$ and $\mathbf{q}_{w}^m$ collect the corresponding values of all nodes.

Now, consider the following modifications of problem \eqref{eqn:mainQP}:
\begin{enumerate}
\item Assume $C(P_0^m)$ to be strictly increasing in $P_0^m$;
\item remove the upper bounds on the current magnitudes;
\item consider shunt capacitors to be modeled by fixed reactive power injections and independent of voltage magnitudes;
\item set $K_{\mathrm{Loss}}=0$; and
\item enforce the constraint $\breve{v}_i^m(p_{c_i},q_{w_i}^m) \le (1+\epsilon) ^2v_0$, instead of $v_i^m \le (1+\epsilon)^2 v_0^2$. 
\end{enumerate}
In addition, following  \cite{GaLiToLo2015}, define for every scenario $m$, $$\underline{A}_i^m:= I + \frac{2}{(1-\epsilon)^2v_0} \begin{bmatrix} r_i \\ x_i \end{bmatrix} \begin{bmatrix} \breve{P}_i^{m-} (\mathbf{p}_{c}^{\min}) & \breve{Q}_i^{m-} (\mathbf{p}_{c}^{\min}, \mathbf{s}_{w}) \end{bmatrix}$$ where $a^-=\min \{a,0\}$.  Also $\mathbf{p}_{c}^{\min}$ and $\mathbf{s}_{w}$ collect all the corresponding values per node.
Then, the main result is that the modified SOCP problem is exact if the following condition holds for every scenario $m$:
\begin{IEEEeqnarray}{rCl}
 \prod\limits_{j=s+1}^{t-1} \underline{A}_{d_j}^m  \begin{bmatrix} r_{d_t} \\ x_{d_t} \end{bmatrix} > 0, \text{for }  2 \le t \le n_d,  0 \le s \le t-2 \label{eqn:sufficient}
\end{IEEEeqnarray}
where $n_d=|\mathcal{P}_{i}|$ for all leaf nodes $i$ (i.e., nodes such that $\mathcal{C}_i=\emptyset$) and $d_j \in \mathcal{P}_{i}$ . 

A sketch of proof based on \cite{GaLiToLo2015} is presented next.  For given $p_{c_i}$ and $q_{w_i}^m$,  we can establish that if $P_i^m$, $Q_i^m$ and $v_i^m$ satisfy \eqref{eqn:repofl}--\eqref{eqn:vopofl}, then the following holds (equivalent to \cite[Lemma 1]{GaLiToLo2015}):
\begin{align}
\breve{P}_i^m(\mathbf{p}_{c})  \le  P_i^m \\
\breve{Q}_i^m(\mathbf{p}_{c}, \mathbf{q}_{w}^m) \le Q_i^m \\
\breve{v}_i^m(\mathbf{p}_{c},\mathbf{q}_{w}^m) \ge v_i^m. 
\end{align}
  Assume we solve problem \eqref{eqn:mainQP} and it turns out that \eqref{eqn:sopofl} is not satisfied with equality in at least one node and one scenario. Here we show that we can construct a feasible solution with a lower objective value.  Call that scenario $m$ and label the node $K$.  Further, without loss of generality, we can assume that the node $K$ is the $k+1$'th node in $\mathcal{P}_{d_n}$ where $d_n$ is a leaf node and $|\mathcal{P}_{d_n}|=n+1$.  Path $\mathcal{P}_{d_n}$ is illustrated as follows: 
\begin{align}
0 \xrightarrow{1}d_1 \xrightarrow{2} d_2 \ldots \xrightarrow{k} d_k=K\xrightarrow{k+1} \ldots \xrightarrow{n} d_n. 
\end{align}
The current solution will be called $s=(\mathbf{P}, \mathbf{Q}, \mathbf{v}, \mathbf{l}, \mathbf{p}_{c}, \mathbf{q}_{w})$, where $\mathbf{P}, \mathbf{Q}, \mathbf{v},\mathbf{l},\mathbf{p}_{c}, \mathbf{q}_{w}$ are vectors collecting all the corresponding values per node and per scenario.  Solution $s$ has the the following property: 
\begin{align}
\frac{(P_K^m)^2 + (Q_K^m)^2}{v_K^m} & < l_K^m, \\
\frac{(P_{d_i}^m)^2 + (Q_{d_i}^m)^2}{v_{d_i}^m} &= l_{d_i}^m \quad \text{ for } i=1,\ldots, k-1.
\end{align}

Algorithm \ref{alg:2} constructs a new feasible solution $s'=(\mathbf{P}',\mathbf{Q}',\mathbf{v}',\mathbf{l}',\mathbf{p}_{c}',\mathbf{q}_{w}')$, which will be proved to have a lower objective. In particular, the new solution $s'$ has the following properties:
\begin{align}
&l_{K}^{'m} < l_{K}^m \Rightarrow \Delta l_{K}^m= l_{K}^{'m}- l_{K}^m < 0, \\
&\frac{ (P_i^{'m})^2 +(Q_i^{'m})^2}{v_i^m} \le l_i^{'m} \text{ for all } i \in \mathcal{N} \backslash \{0\} \label{eqn:newfeasibleproperty2}.
\end{align}

\begin{algorithm}
\caption{Constructing $s'$ from $s$ with lower objective }
\label{alg:2}
\begin{algorithmic}[1]
\State   Initialization: $s' \leftarrow s$ ,  $ v'_0 \leftarrow v_0$ , $\mathcal{N}_{\mathrm{visit}}=\{0\}$ .
\State  Backward sweep:  \textbf{For} $i=k,\ldots,1$ \textbf{do} 
\begin{IEEEeqnarray}{rCl} 
&l_{d_i}^{'m} &\leftarrow \frac{(P_{d_i}^{'m})^2+(Q_{d_i}^{'m})^2}{v_{d_i}^m} \label{eqn:l'} \\
&P_{d_{i-1}}^{'m} &\leftarrow \sum\limits_{j \in \mathcal{C}_{d_{i-1}}} P_{j}^{'m}+r_{j}l_{j}^{'m}+P_{L_{d_i}}^m\notag \\ && + \: p'_{c_{d_i}}-w_{d_i}^m \\
&Q_{d_{i-1}}^{'m} &\leftarrow \sum\limits_{j \in \mathcal{C}_{d_{i-1}}} Q_{j}^{'m} + x_{j}^{m}l_{j}^{'m}+ Q_{L_{d_i}}^m \notag \\ &&+ \: \left(\sqrt{\frac{1}{\mathrm{PF}_{d_i}^2}-1}\right) p'_{c_{d_i}}-q_{w_{d_i}}^{'m}.
\end{IEEEeqnarray}
\State  Forward sweep: 
\textbf{While} $\mathcal{N}_{\mathrm{visit}} \neq \mathcal{N}$ \textbf{do} 
\begin{align}
&\text{find } j \notin \mathcal{N}_{\mathrm{visit}},  i \in \mathcal{N}_{\mathrm{visit}} \text{ such that } j \in \mathcal{C}_i,\\
&v_{j}^{'m}=v_{i}^{'m}-2r_iP_{j}^{'m}-2x_iQ_{j}^{'m}-(r_i^2+x_i^2)l_{j}^{'m}, \\
&\mathcal{N}_{\mathrm{visit}} \leftarrow \mathcal{N}_{\mathrm{visit}} \cup \{j\}. 
\end{align}
\end{algorithmic}
\end{algorithm}

At this point, by proving that $v_i^{'m} \ge v_i^m$, we can use \eqref{eqn:newfeasibleproperty2} to show that $s'$ is feasible.   Furthermore, by additionally proving that $P_0^{'m} < P_0^{m}$, the new solution will have a smaller objective value, which yields a contradiction.  These two facts are proved next.

 Define $\Delta P_i^m = P_i^{'m} - P_i^{m}$ and $\Delta Q_i^m = Q_i^{'m} - Q_i^{m}$.  We can establish the following on path $\mathcal{P}_{d_k}$:
\begin{align}
\begin{bmatrix}
\Delta P_{d_{i-1}}^{m} \\
\Delta Q_{d_{i-1}}^{m}
\end{bmatrix} = B_{d_i}^m\begin{bmatrix} \Delta P_{d_i}^m \\ \Delta Q_{d_i}^m \end{bmatrix}
\end{align}
where for  $i=k, \ldots, 1$:
 \begin{align}
B_{d_i}^m=( I + \frac{2}{v_{d_i}^m} \begin{bmatrix} r_{d_i} \\ x_{d_i} \end{bmatrix} \begin{bmatrix} \frac{P_{d_i}^{'m}+P_{d_i}^m}{2} & \frac{Q_{d_i}^{'m}+ Q_{d_i}^m}{2} \end{bmatrix} )
\end{align}
and 
\begin{align}
\begin{bmatrix} \Delta P_{d_k}^m \\ \Delta Q_{d_k}^m\end{bmatrix}=  \begin{bmatrix} r_{d_k} \\ x_{d_k} \end{bmatrix} \Delta l_K^m.
\end{align}
Thus we can  write for $s=k-2, \ldots, 0$ 
\begin{align}
\begin{bmatrix}
\Delta P_{d_s}^{m} \\
\Delta Q_{d_s}^{m}
\end{bmatrix}= \prod\limits_{i=s+1}^{k-1} B_{d_i}^m  \begin{bmatrix} r_{d_k} \\ x_{d_k} \end{bmatrix} \Delta l_{d_k}^m.
\end{align}
Observe that $B_{d_i}^m - \underline{A}_{d_i}^m = \begin{bmatrix} r_{d_i} \\ x_{d_i} \end{bmatrix} b_{d_i}^T$, where \begin{align}
b_{d_i}=\begin{bmatrix}\frac{P_{d_i}^{'m}+P_{d_i}^m}{2v_{d_i}^m} -\frac{\breve{P}_{d_i}^{m-} (\mathbf{p}_{c}^{\min})}{(1-\epsilon)^2v_0}  \\ \frac{Q_{d_i}^{'m}+ Q_{d_i}^m}{2v_{d_i}^m}-\frac{\breve{Q}_{d_i}^{m-} (\mathbf{p}_{c}^{\min}, \mathbf{s}_w)}{(1-\epsilon)^2v_0} \end{bmatrix} \ge 0,
\end{align}
for $i=k-1, \ldots, 1$. Therefore, \cite[Lemma 3]{GaLiToLo2015} can be used to show that  $P_{d_i}^{'m} < P_{d_i}^m$ and $Q_{d_i}^{'m} < Q_{d_i}^m$ for $i=k-1, \ldots,0$. Notice that $d_0=0$ and hence the following holds: 
\begin{align}
 P_0^{'m} < P_0^m \label{eqn:p0p0prime}.
 \end{align}

Next, we show that $v_i^{'m} \ge v_i^m$. Define $\Delta v_i^m = v_i^m-v_i^{'m}$.  For $A_i \notin \mathcal{P}_{K}$, we have that 
\begin{align}
\Delta v_{i}^m-\Delta v_{A_i}^m = -2 r_i \Delta P_i^m- 2 x_i \Delta Q_i^m -(r_i^2+x_i^2) \Delta l_i^m=0.
\end{align}
For $A_i \in \mathcal{P}_K$, it holds that
\begin{align} 
\Delta v_{i}^m-\Delta v_{A_i}^m =- 2 r_i \Delta P_i^m- 2 x_i \Delta Q_i^m - (r_i^2+x_i^2) \Delta l_i^m \ge 0. 
\end{align}
Adding these inequalities over path $\mathcal{P}_{d_k}$ yields
\begin{align}
\Delta v_{i}^m - \Delta v_{0} \ge 0 \Rightarrow \Delta v_{i}^m \ge 0, \text{ for } i \in \mathcal{P}_{d_k},
\end{align}
which proves that $v_i^{'m} \ge v_i^m$.
We have shown that solution $s'$ is feasible.  Due to \eqref{eqn:p0p0prime} and the assumption that $C(P_0)$ is strictly increasing, the new solution $s'$ has a smaller objective; this is a contradiction.

We have derived sufficient condition  \eqref{eqn:sufficient} under which the SOCP relaxation is exact for a modified problem close to problem \eqref{eqn:mainQP}.  This sufficient condition can be checked a priori.  Notice that modifications 1--3 and 5 are exactly the same as the ones proposed in \cite{GaLiToLo2015} where the problem is not stochastic.

Condition \eqref{eqn:sufficient} is stated per scenario.  It is also possible to state a single sufficient condition that does not depend on $m$.  Specifically, replace $w_i^m$'s with $\max_{m} w_i^m$ in \eqref{eqn:phatpcipl}; then, the resulting matrix $\underline{A}_i$ does not depend on $m$, and condition \eqref{eqn:sufficient} is stated with $\underline{A}_i$ instead of $\underline{A}_i^m$.  This latter condition is a more stringent sufficient condition that accounts for the lowest-consumptions.  We have numerically verified that this latter sufficient condition holds for the network in the numerical tests.

\section{Closed-form updates for $\tilde{P}_i^m,\tilde{Q}_i^m, \tilde{v}_i^m, \tilde{l}_i^m$}
\label{sec:appendixc}
Let $z_1=\tilde{P}_i^m$, $z_2=\tilde{Q}_i^m$, $z_3=\sqrt{\frac{|\mathcal{C}_i|+1}{2}}\tilde{v}_i^m$ and $z_4=\tilde{l}_i^m$.  The z-update for these variables will be equivalent to solving the following optimization problem: 
\begin{subequations}
\label{eqn:pqvlupdate}
\begin{IEEEeqnarray}{rl}
&\min_{\substack{z_1,z_2,z_3,z_4}} \sum\limits_{i=1}^4 (z_i^2+c_iz_i) \\
&\text{subject \ to}  \\
&z_3^{\min} \le z_3 \le z_3^{\max} \label{eqn:z3bounds}\\
&\frac{z_1^2+z_2^2}{z_3} \le k^2z_4 \label{eqn:socpbound} \\
&z_4 \le z_4^{\max} \label{eqn:z4bound}
\end{IEEEeqnarray}
\end{subequations}

where
\begin{IEEEeqnarray*}{lll}
k^2=\sqrt{\frac{2}{|\mathcal{C}_i|+1}} \\
z_3^{\max}=\sqrt{\frac{|\mathcal{C}_i|+1}{2}}(1+\epsilon)^2v_0 \\
z_3^{\min}=\sqrt{\frac{|\mathcal{C}_i|+1}{2}}(1-\epsilon)^2v_0 \\
c_1=-(P_i^m+\hat{P}_i^m+\frac{\lambda_i^m+\hat{\lambda}_i^m}{\rho}) \\
c_2=-(Q_i^m+\hat{Q}_i^m+\frac{\mu_i^m+\hat{\mu}_i^m}{\rho})\\
c_3=-(v_i^m+\sum\limits_{j \in \mathcal{C}_i}\hat{v}_j^m+\frac{\gamma_i^m+\hat{\gamma}_i^m}{\rho} )\\
c_4=-(l_i^m+\hat{l}_i^m+\frac{\gamma_i^m+\hat{\gamma}_i^m}{\rho}). 
\end{IEEEeqnarray*}
Problem \eqref{eqn:pqvlupdate} without considering constraint \eqref{eqn:z4bound} is solved in closed-form in \cite[Appendix I]{PeLo14}.  Using a similar approach, we develop a methodology to obtain a closed-form solution to problem \eqref{eqn:pqvlupdate}, when it includes constraint \eqref{eqn:z4bound}.

Let $\bar{\lambda}, \underline{\lambda}, \mu$ and $\gamma \ge 0$ be Lagrange multipliers corresponding to  \eqref{eqn:z3bounds}, \eqref{eqn:socpbound} and \eqref{eqn:z4bound} respectively.  The KKT conditions for problem \eqref{eqn:pqvlupdate} are: 
\begin{subequations}
\label{eqn:KKT}
\begin{IEEEeqnarray}{lll}
2z_1+c_1+2\mu \frac{z_1}{z_3}=0 \\
2z_2+c_2+2\mu\frac{z_2}{z_3}=0 \\
2z_3+c_3-\mu \frac{z_1^2+z_2^2}{z_3^2}+\bar{\lambda}- \underline{\lambda}=0 \\
2z_4+c_4-k^2\mu+ \gamma =0 \\
\bar{\lambda}(z_3-z_3^{\max})=0 \\
\underline{\lambda}(z_3^{\min}-z_3)=0 \\
\mu(\frac{z_1^2+z_2^2}{z_3}-k^2z_4)=0 \\
\gamma(z_4-z_4^{\max})=0 \label{eqn:z4comp}.  
\end{IEEEeqnarray}
\end{subequations}

The closed-form solution for the KKT conditions in \eqref{eqn:KKT} is obtained by enumerating the cases for $\gamma$:

\subsection{Case 1: $\gamma=0$ }
In this case, as detailed in  \cite[Appendix I]{PeLo14},  there exists a unique solution to \eqref{eqn:KKT} which can be obtained in closed form.  If the obtained closed-form solution satisfies $z_4 \le z_4^{\max}$ then it is also optimal for \eqref{eqn:pqvlupdate}. Otherwise, we need to proceed to the next case. 
\subsection{Case 2: $\gamma > 0$}
In this case, using \eqref{eqn:z4comp}, we can establish that $z_4^*=z_4^{\max}$. Now we examine possible choices for $\mu$. 
\subsubsection{$\mu=0$} 
\begin{IEEEeqnarray}{lll}
z_1^*=-\frac{c_1}{2},
z_2^*=-\frac{c_2}{2},
z_3^*=-\big[\frac{c_3}{2}\big]_{z_3^{\min}}^{z_3^{\max}}.  \notag
\end{IEEEeqnarray}

\subsubsection{$\mu >0$, $z_3^{\min} < z_3 < z_3^{\max}$}
\begin{IEEEeqnarray}{lll}
z_3^*=\mathrm{solve}(az_3^2+bz_3+c=0) \notag \\
\text{where} \notag \\
a=k^2z_4^{\max}(2+\frac{4}{z_4^{\max}})^2 \notag \\
b=\frac{4}{z_4^{\max}}c_3(2+\frac{4}{z_4^{\max}})k^2z_4^{\max}-(c_1^2+c_2^2) \notag\\
c=\frac{4k^2}{z_4^{\max}}c_3^2 \notag\\
z_1^*=-\frac{c_1z_3^*}{2z_3^*+\frac{4}{z_4^{\max}}z_3^*+\frac{2}{z_4^{\max}}c_3} \notag \\
z_2^*=\frac{c_2}{c_1}z_1^*.  \notag
\end{IEEEeqnarray}

\subsubsection{$\mu >0$, $z_3^{\min} =z_3$}
\begin{IEEEeqnarray}{lll}
z_3^*=z_3^{\min} \notag \\
z_1^*=-\frac{c_1z_3^*}{2\mu^*+2z_3^*} \notag \\
z_2^*=-\frac{c_2}{c_1}z_1^* \notag \\
\mu^*=\frac{\sqrt{c_1^2+c_2^2}\sqrt{z_3^{\min}}}{2\sqrt{k^2z_4^{\max}}}-z_3^{\min}. \notag
\end{IEEEeqnarray}
\subsubsection{$\mu >0$, $z_3^{\max} =z_3$}
\begin{IEEEeqnarray}{lll}
z_3^*=z_3^{\max} \notag \\
z_1^*=-\frac{c_1z_3^*}{2\mu^*+2z_3^*} \notag \\
z_2^*=-\frac{c_2}{c_1}z_1^*  \notag \\
\mu^*=\frac{\sqrt{c_1^2+c_2^2}\sqrt{z_3^{\max}}}{2\sqrt{k^2z_4^{\max}}}-z_3^{\max}. \notag 
\end{IEEEeqnarray}

\bibliographystyle{IEEEtran}
\bibliography{IEEEabrv,biblio}

\end{document}